\documentclass[runningheads]{llncs}
\usepackage[T1]{fontenc}
\usepackage{tabularx}
\usepackage{booktabs}
\usepackage{makecell}
\usepackage{adjustbox}
\usepackage{array}
\usepackage{multirow}
\usepackage[table]{xcolor}
\usepackage{colortbl}

\newcolumntype{Y}{>{\raggedright\arraybackslash}X}
\usepackage{appendix}
\usepackage{amsmath} 
\usepackage{amssymb}
\usepackage[square,numbers]{natbib}
\usepackage{tikz}

\newtheorem{runningexample}{Running Example}
\usepackage{fontawesome5} % For icons in the Workbench diagram
\tikzset{
  fwBox/.style={draw, rounded corners, align=center, inner sep=6pt, minimum width=28mm, minimum height=12mm},
  fwRole/.style={draw, rounded corners, align=center, inner sep=4pt, minimum width=25mm, minimum height=8mm},
  fwStore/.style={draw, rounded corners, align=center, inner sep=4pt, minimum width=40mm, minimum height=10mm},
  fwQueue/.style={draw, rounded corners, align=center, inner sep=3pt, minimum width=16mm, minimum height=7mm},
  fwArrow/.style={->, line width=0.55pt},
  fwDotted/.style={->, dotted, line width=0.55pt}
}
\usetikzlibrary{positioning,arrows.meta}

\tikzset{
wBox/.style={
    draw,
    rounded corners,
    align=center,
    inner xsep=3pt,
    inner ysep=5pt,
    minimum width=30mm,
    minimum height=10mm
  },
  fwStore/.style={
    draw,
    rounded corners,
    align=center,
    inner xsep=3pt,
    inner ysep=4pt,
    minimum width=50mm,
    minimum height=10mm
  },
  fwArrow/.style={->, line width=0.55pt},
  fwDot/.style={->, dotted, line width=0.55pt}
}

\tikzset{
  oodaPhase/.style={
    draw,
    rounded corners,
    align=center,
    inner xsep=4pt,
    inner ysep=5pt,
    minimum width=24mm,
    minimum height=10mm
  },
  oodaStore/.style={
    draw,
    rounded corners,
    align=center,
    inner xsep=5pt,
    inner ysep=5pt,
    minimum width=42mm,
    minimum height=11mm
  },
  oodaArrow/.style={->, line width=0.55pt}
}
\usetikzlibrary{shapes.geometric, arrows.meta, arrows, positioning, fit, calc, backgrounds}
\usepackage{bbding}
\usepackage{graphicx}
\begin{document}
\title{The Alignment Flywheel: A Governance-Centric Hybrid MAS for Architecture-Agnostic Safety}
\titlerunning{The Alignment Flywheel}
\author{Elias Malomgré\inst{1}\inst{2}\orcidID{0009-0000-3052-7047} \Envelope \and
Pieter Simoens\inst{1}\inst{3}\orcidID{0000-0002-9569-9373}
\institute{IDLab, Ghent University - imec, Belgium \and
\email{elias.malomgre@ugent.be} \and
\email{pieter.simoens@ugent.be}}}
\authorrunning{E. Malomgré et Simoens.}

\maketitle              % typeset the header of the contribution
\begin{abstract}
Multi-agent systems provide mature abstractions for role decomposition, coordination, and normative governance, but increasingly capable learned components make post-deployment safety harder to inspect, audit, and update. When safety behavior is absorbed into a decision component, narrow failures may require retraining or rollback of the full component. This instantiates our vision of the Alignment Flywheel as a governance-centric hybrid MAS architecture that decouples decision generation from safety governance. We denote the agent or policy that generates candidate trajectories as the Proposer; it passes its output to a governed Safety Oracle stack, which returns safety scores, prediction uncertainty, audit coverage uncertainty, and evidence hooks through a stable interface. An Enforcement layer applies explicit risk policy at runtime. Around this loop, a governance MAS performs monitoring, red-teaming, verification, triage, refinement, and versioned release management. The central engineering principle is patch locality: many newly observed safety failures can be mitigated through small governance batches for the Oracle stack and its audit state rather than by retraining or retracting the Proposer. The architecture is implementation-agnostic with respect to both Proposer and Oracle. It defines the roles, artifacts, protocols, and release semantics needed for runtime gating, audit intake, signed updates, staged rollout, and rollback. We demonstrate executability in two scenarios: a learned spatial Oracle patched through regression-checked governance updates, and a clinical GenAI proxy setting illustrating structured norms, escalation, and audit coverage. Our implementation code and documentation are available open source at https://github.com/decide-ugent/Alignment-Flywheel.

\end{abstract}

\section{Introduction}
As models and autonomous components become more capable, a central challenge is no longer only how to make them perform well, but how to make them behave acceptably under the rules, norms, and risk limits of the settings in which they are deployed, often discussed as \emph{alignment}. These failures often emerge at component interfaces, particularly when components run different versions or are updated at different times. In complex ML systems, entangled dependencies and hidden feedback loops can make the resulting regressions difficult to localize \cite{sculley2015hidden,breck2017ml}. In production, such failures are frequently \emph{silent}, becoming tractable only through end-to-end observability across the full pipeline \cite{shankar2021towards}, and these operational governance problems are cross-component and lifecycle-driven rather than localized to model internals \cite{ZAROUR2025107733}.

A recurring difficulty is that rules a system should follow are often absorbed into the internal parameters of the decision policy itself \cite{chaudhari2025rlhf,rafailov2023direct,sun2025inverse}. This applies to any autonomous agent, not only foundation models, whose behavior changes through retraining, optimization, model replacement, or configuration updates. When acceptable behavior is encoded mainly in the policy, it becomes harder to inspect, patch, qualify, and redeploy that behavior when norms, regulations, or risk tolerances change \cite{malomgr2026}, especially under dataset shift and drift \cite{quinonero2008dataset,hovakimyan2024evolving}. In this paper, we use \emph{governance} in an operational sense: a layer that decides whether proposed behavior should be allowed, blocked, deferred, escalated, patched, or rolled back, and how those decisions are updated over time. We, therefore, propose a different engineering unit of change for that layer: small, targeted governance patches that constrain newly unsafe trajectory classes without requiring full policy retraction or redeployment. This resembles runtime enforcement \cite{qin2026governed} and \emph{shielding} \cite{alshiekh2018safe,konighofer2025shields}, but our focus is on the engineering needed to make governance patchable, auditable, and operable across a hybrid MAS deployment.

Building on our Alignment Flywheel vision \cite{malomgr2026} and our prior work shows that IIRL can instantiate a standalone Safety Oracle queried independently of the underlying Proposer policy in robotic locomotion \cite{malomgre2025mixture}; this paper specifies this vision as a hybrid MAS architecture, protocol layer, and deployment semantics needed to operate such an Alignment Flywheel in practice. This is the first step toward making the Alignment Flywheel operational by providing an implementation that could fulfill the specified use cases. Still, we cannot yet show empirical evidence for all of them. The earlier work framed alignment as an ongoing governance process rather than a one-time training outcome; the present paper turns that lifecycle view into an executable Proposer-Oracle governance architecture. IIRL is used as a convenient reference instantiation, but the Oracle role is method-agnostic: any safety artifact satisfying the same interface assumptions can be governed by the architecture.

Concretely, we denote the agent or policy that generates candidate actions or plans as trajectories as a Proposer, and its output as a Safety Oracle that returns safety-relevant signals through a stable interface contract. The decision component does not contain all normative intelligence itself. Instead, an Enforcement layer interprets Oracle outputs under an explicit risk policy, and a governance MAS supervises the Oracle through monitoring, auditing, uncertainty-driven verification, and versioned refinement. The central engineering principle is \emph{patch locality}: many safety fixes for the Proposees are applied to the governed Oracle artifact and its update pipeline rather than by modifying or retracting policy. This is particularly valuable in hybrid systems \cite{bak2014real,phan2017component}, where components evolve at different cadences and incur different qualification costs. The policy upgrades may be expensive to train and certify, whereas Oracle patches can be small, reviewable, and narrowly scoped to newly observed failures at the proposer-oracle boundary \cite{sculley2015hidden,breck2017ml,shankar2021towards}.

From a systems-engineering perspective, the architecture applies separation of governance concerns. Monitoring, escalation, verification, refinement, enforcement, and release management are factored into interacting modules with explicit protocols and stable interface contracts. This supports modular contribution: one can improve a single role in the governance loop without understanding the internal learning dynamics of the Proposer or the full construction of the Oracle, provided the module consumes and produces the agreed artifacts correctly. A central part of this separation is the release model: Oracle updates are treated as versioned governance artifacts that can be rolled out across a fleet, monitored for regressions, and rolled back under bounded propagation delay \cite{shahin2017continuous}. Since these patches function as distributed governance controls, the architecture incorporates signed update metadata \cite{newman2022sigstore}, anti-rollback checks \cite{kuppusamy2017mercury}, and monitoring for proposer-oracle interface regressions such as uncertainty-calibration drift \cite{ovadia2019can} and decision-distribution shifts conditioned on Proposer version \cite{shankar2021towards}.

\begin{runningexample}[Clinical GenAI Assistant]
A hospital deploys a \\ clinician-facing Clinical GenAI Assistant that drafts patient-facing messages, summarizes chart context, retrieves approved institutional guidance, and proposes workflow actions. The Assistant acts as the Proposer. A Safety Oracle evaluates candidate replies or action plans before execution. If the Oracle's prediction is uncertain or if the Alignment Flywheel has insufficient audit coverage for this kind of case, the system must not answer definitively. Instead, it must abstain (e.g., respond with ``I don't know'') or escalate for clinical review.
\end{runningexample}

This paper makes four contributions to the engineering of deployable hybrid multi-agent systems. First, it defines a Proposer-Oracle topology and a trajectory-level gating model that applies to both single-step actions and multi-step plans across domains and modalities. Second, it specifies the Alignment Flywheel as an executable hybrid MAS design, with coordinated roles for monitoring, escalation, auditing, refinement, and enforcement, together with the artifacts they exchange and their authority boundaries. Third, it formalizes an Oracle interface contract covering decision outputs, prediction uncertainty, audit coverage uncertainty, and evidence hooks, enabling audit and patch workflows while preserving architectural invariants such as patch locality and version stability. Fourth, it introduces deployment semantics for proposer-oracle systems in which safety fixes are released as small, versioned Oracle patches rather than as full policy redeployments, including progressive rollout, regression monitoring, rollback under bounded propagation delay, and signed update metadata for fleet distribution.

\section{Background and Positioning}
This section positions the paper relative to operational governance, candidate Safety Oracle constructions, and runtime-control architectures. The goal is not to argue for one universally best Oracle method, but to clarify the interface assumptions that make an artifact governable, patchable, and deployable. A broader discussion of alternative Oracle families is deferred to Appendix~\ref{app:extended-background}.

\paragraph{Governance}
In this paper, \emph{governance} is used in an operational rather than purely prescriptive sense. Much AI governance work specifies principles, documentation requirements, explanations, or organizational responsibilities \cite{mantymaki2022defining, birkstedt2023ai, batool2025ai}; such mechanisms are necessary, but they do not, by themselves, determine how a running autonomous system should react when a new failure class is observed. Here, governance refers to the mechanisms that control proposed behavior and manage corrective updates during deployment. Explicit constraints, uncertainty signals, and verification results inform whether a proposed trajectory is trusted, blocked, deferred, or escalated. Governance also covers how corrective patches are released and rolled back. This scope extends beyond runtime filtering alone \cite{qin2026governed}, while remaining narrower than alignment in its broadest sense. We focus on how governance judgments are represented, exchanged, enforced, and updated across interacting modules in a deployed system. This view is compatible with normative MAS work that treats governance as explicit administration through norms and related state rather than as a property hidden inside individual agents or policies \cite{singh2014norms,boella2006introduction}.

\paragraph{Candidate Safety Artifacts and Oracle Constructions}
A Safety Oracle is any artifact that evaluates candidate actions, trajectories, or plans through a stable interface and returns deployment-relevant signals such as a safety score, uncertainty, or evidence hooks. Several artifact families can instantiate this role; preference-shaped policies such as RLHF and DPO improve behavior directly, but tend to entangle governance with the policy, making independent patching and audit harder \cite{ziegler2019fine,ouyang2022training,rafailov2023direct}; demonstration-derived evaluators, including IRL-style reward models, externalize evaluation but depend on expert coverage and may generalize poorly outside demonstrated regions \cite{ng2000algorithms,piet,ziebart2008maximum,adams2022survey,arora2021survey}; constraint- and logic-based artifacts, guardrails, monitors, and shields can be more interpretable or formally grounded, but require the relevant structure to be specified and observable at runtime \cite{liu2024comprehensive,rebedea2023nemo,alshiekh2018safe,konighofer2025shields}. IIRL appears especially promising because it yields a standalone evaluative artifact with comparatively weak operational assumptions \cite{malomgr2026,malomgre2025mixture}; however, the Proposer--Oracle design remains method-agnostic. Appendix~\ref{app:oracle-families} expands this comparison.

\paragraph{Hybrid Agent Architectures, Interfaces, and Governed Deployment}
The core engineering difficulty in hybrid agent systems is the asynchronous evolution of heterogeneous components. A Proposer may change through retraining or replacement at a slow, expensive cadence. At the same time, governance constraints, guardrails, or shielding mechanisms must adapt much more quickly as new failure cases are observed. This concentrates risk at interfaces, where representational drift, version skew, and calibration mismatch can produce hard-to-localize regressions \cite{sculley2015hidden,breck2017ml}. Pipeline-level observability is therefore needed to attribute failures across data, models, and surrounding components \cite{shankar2021towards}. A practical deployment analog of this external governance is the guardrails pattern: constraining a powerful generator at runtime using modular checks and rules external to the base model, as in NeMo Guardrails \cite{rebedea2023nemo}. Runtime enforcement and shielding similarly separate optimizing policies from external safety mechanisms, often with stronger formal guarantees \cite{alshiekh2018safe,konighofer2025shields}. In the Alignment Flywheel architecture, such guardrails and shields can themselves be treated as candidate safety artifacts, provided they expose the required governance interface.

\section{System Overview: Roles, Artifacts, and Interfaces}
\label{sec:system}

\begin{figure}
    \centering
    \includegraphics[width=1\linewidth]{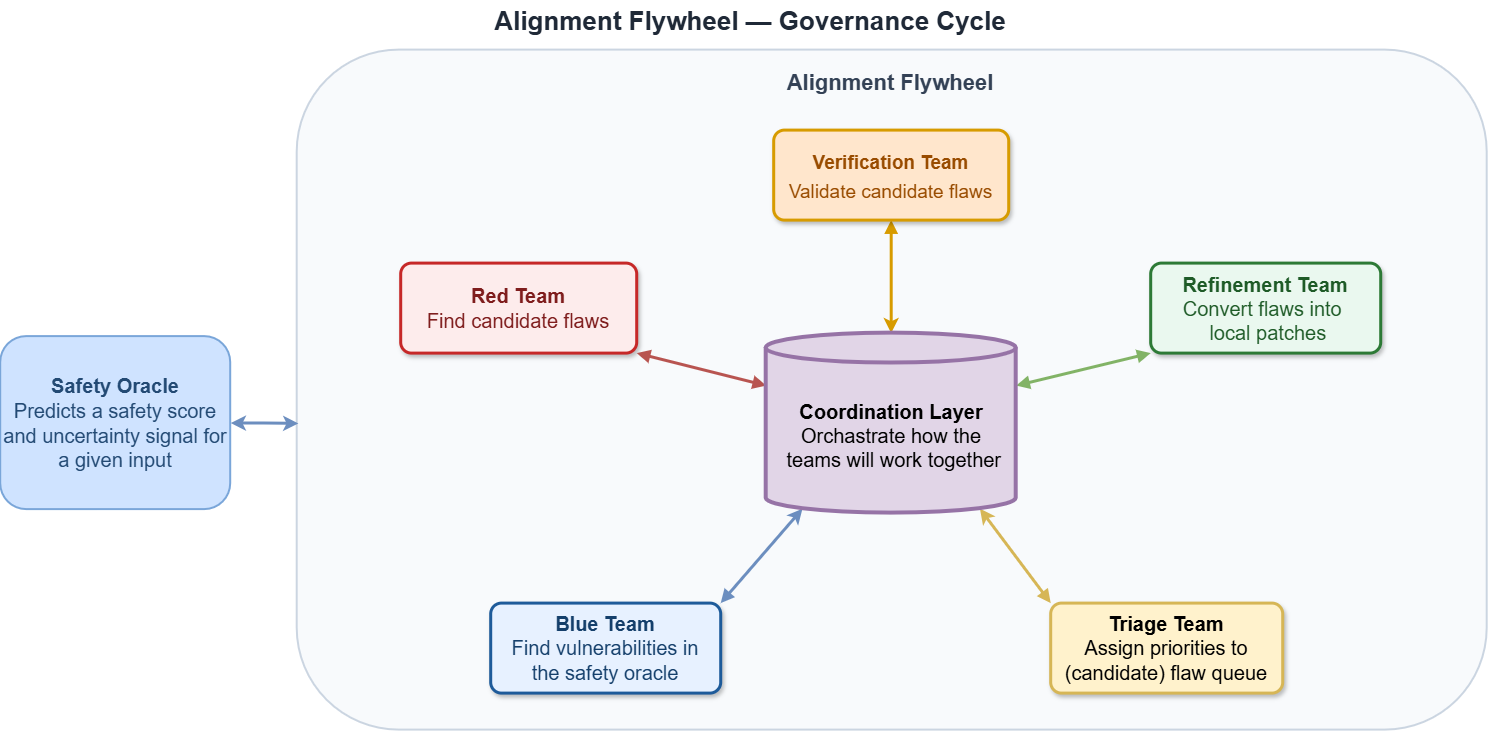}
    \caption{An overview of the Alignment Flywheel and its roles.}
    \label{fig:placeholder}
\end{figure}

The Alignment Flywheel, as shown in Figure \ref{fig:placeholder}, treats AI safety not only as a training problem, but as an externalized control problem. It wraps a Safety Oracle with a governance layer that checks candidate trajectories against explicit norms, operational evidence, prediction uncertainty, and audit coverage. This is especially relevant when a third party supplies the Oracle: the vendor may provide a strong statistical artifact, but cannot be expected to encode dynamic regulations, organization-specific compliance interpretations, approval rules, or documentation practices. The Alignment Flywheel therefore maintains the normative specification $\Phi$ and supervises the Oracle against current governance policy.

The same mechanism supports alignment-as-a-service during training, fine-tuning, or online adaptation. The Alignment Flywheel monitors visited contexts and trajectory segments, routing uncertain or insufficiently audited regions to verification and refinement. It uses end-to-end observability to attribute failures to cross-component changes rather than isolated modules \cite{shankar2021towards}. Because verification and refinement capacity may be limited, oversight is designed to be scalable: agents handle routine discovery, clustering, and patch preparation, while humans can intervene at the level of norms, thresholds, escalation rules, and release approvals. Appendix~\ref{app:oversight} details this tunable oversight model. In this sense, the Alignment Flywheel acts as an independent regulatory shell for runtime enforcement, audit intake, refinement, and documentation support \cite{raji2022outsider}. The minimal notation needed to follow the rest of the paper is shown in Table~\ref{tab:notation}.

\begin{table}[h]
\centering
\small
\caption{Minimal notation used in the architecture.}
\begin{tabular}{ll}
\hline
Symbol & Meaning \\
\hline
$\Sigma$ & context (task state, inputs, metadata; any modality) \\
$\tau$ & trajectory (action, tool call, message, or plan) \\
$P$ & Proposer producing $\tau_{\mathrm{cand}}$ from $\Sigma$ \\
$O$ & Safety Oracle (third-party statistical artifact) \\
$s$ & Safety Oracle raw safety score \\
$u$ & Safety Oracle prediction uncertainty \\
$u_{thresh}$ & vendor-defined threshold for acceptable prediction uncertainty \\
$u_a$ & audit coverage uncertainty generated by the Alignment Flywheel \\
$u_{a,thresh}$ & threshold for acceptable audit coverage uncertainty \\
$\Delta O$ & local correction inside a governance batch \\
$B_O$ & batched governance update for a Safety Oracle \\
$E$ & Enforcement Layer (gates execution, logs evidence) \\
$\Phi$ & Normative Specification (maintained by the MAS) \\
$v_O$ & Oracle / governance-batch version identifier \\
$C$ & Coordination layer \\
$K$ & Knowledge Base (demonstrations, failures, patches, ledger) \\
$Q_{ver}$ & Verification Queue (raw candidates from Red Team / monitoring) \\
$Q_{ref}$ & Refinement Queue (clustered / prioritized flaws for correction) \\
\hline
\end{tabular}
\label{tab:notation}
\end{table}

\subsection{Multi-Agent Governance Roles}
\label{sec:roles}
The architecture comprises five specialized roles as shown in Table \ref{tab:roles} and Figure \ref{fig:placeholder}. Crucially, each role can be instantiated by an autonomous agent, a human expert, or a hybrid collective, supporting configurable degrees of autonomy and escalation \cite{scerri2001adjustable}.

\begin{table}[h!]
\centering
\small
\caption{Governance roles in the Alignment Flywheel.}
\setlength{\tabcolsep}{3pt}
\renewcommand{\arraystretch}{1.08}
\begin{adjustbox}{max width=\columnwidth}
\begin{tabularx}{\columnwidth}{@{}lYYY@{}}
\arrayrulecolor{black}
\toprule
\textbf{Role} & \textbf{Goal} & \textbf{Input} & \textbf{Output} \\
\midrule

Red Team & find false negatives & $O,\tau,\Sigma$ & candidates for $Q_{ver}$ \\
\arrayrulecolor{black!50}\specialrule{0.3pt}{0pt}{0pt}\arrayrulecolor{black}

\multirow{4}{*}{Blue Team}
  & detect drift & traces, stats & drift alerts \\
    & detect vulnerabilities & failures, anomalies & audit intake \\
  & monitor deployment & runtime traces & deployment reports \\
  & steer Red Team & risk summaries & audit priorities \\
\arrayrulecolor{black!50}\specialrule{0.3pt}{0pt}{0pt}\arrayrulecolor{black}

Verification Team & check against $\Phi$ & $Q_{ver}$ items & verified breaches \\
\arrayrulecolor{black!50}\specialrule{0.3pt}{0pt}{0pt}\arrayrulecolor{black}

\multirow{2}{*}{Triage Team}
  & govern $Q_{ver}$ & candidate cases & prioritized $Q_{ver}$ \\
  & govern $Q_{ref}$ & verified breaches & prioritized $Q_{ref}$ \\
\arrayrulecolor{black!50}\specialrule{0.3pt}{0pt}{0pt}\arrayrulecolor{black}

Refinement Team & patch and package & $Q_{ref}$ items & $\Delta O$, $B_O$ \\
\arrayrulecolor{black}
\bottomrule
\end{tabularx}
\end{adjustbox}
\label{tab:roles}
\end{table}

\subsection{Coordination Layer $C$}
\label{sec:kb}
The Alignment Flywheel does not require a particular coordination mechanism. We specify a coordination layer $C$ that provides the semantic operations needed by the governance roles: publishing typed governance records, routing work items to role-specific backlogs, preserving lineage between records, exposing queryable operational views, and recording versioned release transitions. The architecture therefore constrains the protocol semantics between roles, not the concrete coordination. In the remainder of the paper, for clarity and traceability, we instantiate $C$ as an append-only blackboard-style knowledge base that stores the artifacts shown in Table \ref{tab:artifacts}; governance-relevant state changes are recorded as immutable, typed artifacts, and operational state is derived by replaying or querying these artifacts rather than by mutating shared in-memory state. This supports stateless agents (Sec.~\ref{sec:ooda}) and makes audits, rollback, and release analysis traceable.

Each artifact in $K$ carries explicit type and lineage information, so derived artifacts record their inputs (e.g., VerificationResult$\rightarrow$CandidateFlaw, \\ GovernanceBatch$\rightarrow$VerifiedBreach set). This supports provenance queries such as ``why was this blocked?'' and lets queues store references into $K$ rather than full payloads.To support monitoring and steering at scale, $K$ also maintains queryable operational views, including norm coverage over $\Phi$, drift and novelty summaries, open breach clusters with risk scores, and fleet rollout state derived from the release ledger.

\begin{table}[t]
\centering
\small
\caption{Typed artifacts stored in $K$.}

\setlength{\tabcolsep}{3pt}
\renewcommand{\arraystretch}{1.05}
\begin{adjustbox}{max width=\columnwidth}
\begin{tabularx}{\columnwidth}{@{}lYY@{}}
\toprule
\textbf{Artifact} & \textbf{Purpose} & \textbf{Producer} \\
\midrule
$\Phi$ & norms, tests, severities & governance authority \\
$F$ & confirmed breaches + evidence & verification \\
$Q_{ver},Q_{ref}$ & queue pointers + priority & monitoring / triage team\\
$B_O$ & batched governance update & refinement team \\
$L$ & rollout / rollback history & release manager \\
reports & drift, vulnerabilities, coverage & blue team  \\
\bottomrule
\end{tabularx}
\end{adjustbox}
\label{tab:artifacts}
\end{table}

\subsection{Oracle Interface Contract (the query interface)}
\label{sec:oracle-contract}
To preserve strict separation of concerns, the third-party Safety Oracle $O$ is treated as a black-box statistical evaluator. It contains no symbolic business logic, nor does it know the current regulations $\Phi$. It evaluates a trajectory and outputs its internal statistical state. The Enforcement Layer $E$ interacts with the deployed oracle stack through a single query interface:
\begin{center}
\fbox{
\begin{minipage}{0.94\linewidth}
\small
\textbf{Inputs:} Context $\Sigma$, candidate trajectory $\tau$, and return flags
$f_s, f_u, f_{u_a}, f_{thresh} \in \{0,1\}$ indicating whether the caller requests the raw safety score, prediction uncertainty, audit coverage uncertainty, and threshold values. \\
\textbf{Outputs (as requested by flags):} \\
(1) $s \in \mathbb{R}$: the raw safety score. \\
(2) $u \in \mathbb{R}$: the Oracle's internal prediction uncertainty for $s$. \\
(3) $u_a \in \mathbb{R}$: the audit coverage uncertainty measured by the Alignment Flywheel. \\
(4) $(u_{thresh}, u_{a,thresh})$: the configured thresholds for acceptable prediction uncertainty and audit coverage uncertainty. \\
(5) $v_O$: the version identifier of the deployed oracle / governance-batch state used.
\end{minipage}}
\end{center}

The query interface is therefore vendor-backed but governance-mediated: vendor-side signals such as $s$ and $u$ are returned by the Safety Oracle itself, while governance-side signals such as $u_a$ are attached from the currently deployed Alignment Flywheel governance state. This preserves a single enforcement-facing protocol while keeping the provenance of the signals distinct.

Rather than distributing individual local corrections or isolated audit-coverage updates, the Flywheel releases signed governance batches $B_O$ as the engineering unit of change. A batch may contain one or more local corrections $\Delta O$, updates to the state used to compute $u_a$, threshold changes, and the linked regression evidence required for rollout and rollback.

\subsection{Enforcement Logic and Prioritization}
\label{sec:enforcement_logic}
Because the Oracle outputs only raw signals, the normative intelligence resides within the MAS. As shown in Figure~\ref{fig:runtime-enforcement}, the Oracle does not decide policy; instead, the Enforcement layer interprets the returned signals, applies the configured risk policy, logs the outcome, and escalates relevant cases for later hardening. A simple risk policy is shown in Table \ref{tab:enforcement}.

\begin{table}[t]
\centering
\small
\caption{Example enforcement policy.}
\setlength{\tabcolsep}{3pt}
\renewcommand{\arraystretch}{1.05}
\begin{adjustbox}{max width=\columnwidth}
\begin{tabularx}{\columnwidth}{@{}YY@{}}
\toprule
\textbf{Condition} & \textbf{Action} \\
\midrule
$u \geq u_{thresh}$ or $u_a \geq u_{a,thresh}$ & fail-closed or halted; send (priority) audit to $Q_{ver}$  \\
$s$ unsafe, uncertainty acceptable & block \\
$s$ safe, uncertainty acceptable & allow; may later be audited \\
\bottomrule
\end{tabularx}
\end{adjustbox}
\label{tab:enforcement}
\end{table}

\begin{figure}[t]
\centering
\begin{tikzpicture}[>=Latex, node distance=10mm and 14mm, every node/.style={font=\footnotesize}]

\node[fwBox] (P) {Proposer $P$\\[-1pt]\scriptsize $\Sigma \mapsto \tau_{\mathrm{cand}}$};
\node[fwBox, right=8mm of P] (E) {Enforcement $E$\\[-1pt]\scriptsize derive $a,u_a$\\[-1pt]\scriptsize log + gate};
\node[fwBox, right=30mm of E] (O) {Safety Oracle $O$\\[-1pt]\scriptsize vendor API};

\draw[fwArrow] (P) -- node[above] {$\tau_{\mathrm{cand}}$} (E);
\draw[fwArrow] (E) -- node[above] {query $(\Sigma,\tau)$} (O);
\draw[fwArrow] (O) -- node[below, align=center]
{\scriptsize raw: $(s,u,u_{thresh},v_O)$\\[-1pt]\scriptsize (opt: $\phi_{hint}$, evid)} (E);

% optional revision hint back to proposer
\draw[fwDot] (E.north) -- ++(0,2mm) -| node[above] {\scriptsize revise($\phi_{hint}$)} (P.north);

\node[fwStore, below=8mm of E] (K) {Knowledge Base $K$ (append-only)\\[-1pt]\scriptsize decision records + evidence + audit intake};

% runtime logging / escalation
\draw[fwDot] (E) -- node[right] {\scriptsize decision record / escalate / audit intake} (K);

\end{tikzpicture}
\caption{Runtime enforcement during deployment. From context $\Sigma$, proposer $P$ generates a candidate trajectory $\tau_{\mathrm{cand}}$. Enforcement $E$ queries the deployed oracle stack, which returns the requested signals, e.g.\ $(s,u,u_{thresh},u_au_{a,thresh},v_O)$, combining vendor-side outputs and Alignment Flywheel-side governance outputs under a single query interface. Enforcement then derives the action $a$ and the uncertainty state $u$, logs the decision, and writes the audit intake to the append-only knowledge base $K$. Dotted paths indicate optional revision and escalation under the configured risk policy.}
\label{fig:runtime-enforcement}
\end{figure}

\begin{runningexample}[Clinical GenAI Assistant: governed query]
A clinician asks the Clinical GenAI Assistant to draft a patient-facing reply about whether a recently prescribed medication should still be taken after the patient reports a possible side effect. The Proposer generates a candidate message and a short action plan. The deployed oracle stack is queried with $(\Sigma,\tau)$, where $\Sigma$ contains the clinical context and retrieved evidence, and $\tau$ is the proposed reply-and-action trajectory. The vendor Oracle returns a safety score $s$ and prediction uncertainty $u$. The Alignment Flywheel layer attaches audit coverage uncertainty $u_a$, reflecting that this class of advice has only limited prior verification coverage under the hospital's current governance policy. Because $u_a \geq u_{a,thresh}$, Enforcement does not allow direct execution and instead routes the case to $Q_{ver}$ for verification.
\end{runningexample}

\section{Interaction Protocols and OODA Dynamics}
\label{sec:interaction}

The Alignment Flywheel coordinates distributed governance agents through role-specific OODA loops (Observe--Orient--Decide--Act) \cite{brehmer2005ooda}. The OODA abstraction separates the \emph{mechanism} of governance, which records are read and written, from the \emph{strategy} used by a role, such as fuzzing, gradient-based search, prompt mutation, symbolic checking, or patch synthesis \cite{li2018fuzzing,ji2023ai,li2024multitude}. Governance agents interact only through the append-only Knowledge Base $K$ and its queues, so agents can be restarted, replicated, or replaced without invalidating global governance state. Appendix~\ref{app:ooda} gives role-specific OODA pseudocode.

\subsection{OODA Abstraction}
\label{sec:ooda}
Each governance role executes its OODA loop over the coordination layer $C$. The role observes governance state through $C$, orients relative to its objective, selects a strategy module, and publishes the resulting governance records back through $C$. In the reference instantiation summarized in Table~\ref{tab:ooda-template}, these operations correspond to reading from $K$ and $Q_{ver}$/$Q_{ref}$ and appending new records to $K$. This keeps the paper’s running example concrete while preserving the architectural distinction between the logical governance protocol and the concrete blackboard-style coordination mechanism.

\begin{table}[h!]
\centering
\small
\caption{Role-independent OODA template for governance agents.}
\label{tab:ooda-template}
\begin{tabular}{p{0.16\linewidth}p{0.39\linewidth}p{0.35\linewidth}}
\hline
\textbf{Phase} & \textbf{Role-independent function} & \textbf{Example} \\
\hline
Observe & Read relevant records from $K$ and queue heads. & Red Team reads active $v_O$ and high-severity norms \\
Orient & Interpret records relative to the role objective. & Red Team identifies a norm with low coverage. \\
Decide & Select a strategy module without changing the protocol. & Prompt mutation to verify norms for an LLM proposer \\
Act & Execute the strategy and write typed records back to $K$. & Candidate flaws are pushed to the queue \\
\hline
\end{tabular}
\end{table}

\begin{runningexample}[Clinical GenAI Assistant]
Red Team agents generate patient-message variants involving medication side effects, missing evidence, or unsafe disposition choices. Low-uncertainty claimed-safe replies are inserted into $Q_{ver}$, verified against $\Phi$, and converted into breach clusters such as ``unsupported medication advice'' or ``under-escalated urgent symptoms''. These clusters become refinement jobs in $Q_{ref}$ and later governed updates.
\end{runningexample}

\subsection{Double-Filter Pipeline}
\label{sec:pipeline}

Coordination is organized as a double-filter pipeline that separates discovery from correction as shown in Table \ref{tab:double-filter}. Stage~1 converts raw candidate flaws into verified breaches; Stage~2 converts verified breaches into governed updates. The two-stage structure prevents high-volume automated probes from overwhelming verification and refinement capacity. Triage can prioritize candidates and breach clusters using risk signals such as norm severity, dangerous certainty $(u_{\mathrm{thresh}}-u)$, novelty relative to prior history in $K$, and operational urgency \cite{zhao2025triage}.

\begin{table}[h!]
\centering
\small
\caption{Double-filter pipeline from discovery to governed update.}
\label{tab:double-filter}
\begin{tabular}{p{0.14\linewidth}p{0.20\linewidth}p{0.42\linewidth}p{0.16\linewidth}}
\hline
\textbf{Stage} & \textbf{Step} & \textbf{Function} & \textbf{Output} \\
\hline
Verification & Injection & Red Team generates candidate trajectories $\tau$ and pushes them to $Q_{ver}$. & CandidateFlaw \\
Verification & Triage Candidates & They are prioritized by $\Delta = u_{thresh}-u$. Low-uncertainty claimed-safe cases are high-value false-negative audit targets. & prioritized $Q_{ver}$ \\
Verification & Validation & Verification checks candidates against $\Phi$, and violations become immutable failure records in $K$. & VerifiedBreach \\
\hline
Refinement & Ingestion & Verified breaches are hash-stored in $K$ for tracing and de-duplication. & breach records \\
Refinement & Triage II & Breaches are clustered into failure families, reducing cognitive load and enabling batch remediation \cite{zhao2025triage,turcotte2025automated}. & breach clusters \\
Refinement & Job creation & Triage packages a prioritized cluster as a repair task. & RefinementJob \\
Refinement & Correction & Refinement synthesizes a patch $\Delta O$ or governance-batch update and commits it to the release ledger $L$. & PatchCommit \\
\hline
\end{tabular}
\end{table}

\subsection{Message Protocols}
\label{sec:protocols}

Inter-agent coordination is enforced through typed records stored in $K$. These records define the handoff points between governance roles and retain enough context for audit, replay, and patch attribution. Table~\ref{tab:message-protocols} summarizes the main records; complete schemas are given in Appendix~\ref{app:protocols}.

\begin{table}[h!]
\centering
\small
\caption{Core protocol records exchanged through $K$ and the queues.}
\label{tab:message-protocols}
\begin{tabular}{p{0.24\linewidth}p{0.22\linewidth}p{0.43\linewidth}}
\hline
\textbf{Record} & \textbf{Producer / route} & \textbf{Purpose} \\
\hline
CandidateFlaw & Red Team $\rightarrow Q_{ver}$ & Discovery intake record containing $(\Sigma,\tau)$ and the Oracle signals observed at discovery time. \\
VerificationResult & Verification $\rightarrow K$ & Explicit governance judgment separating raw suspicion from a confirmed or rejected violation of $\Phi$. \\
VerifiedBreach & Verification $\rightarrow K$ & Durable failure record created for confirmed violations; later clustered, prioritized, and linked to patches. \\
RefinementJob & Triage $\rightarrow Q_{ref}$ & Clustered repair task that converts many verified breaches into one prioritized refinement unit. \\
PatchCommit & Refinement $\rightarrow K,L$ & Governed update record binding a patch to its parent version, motivating breaches, regression evidence, and authorization signature. \\
\hline
\end{tabular}
\end{table}

\section{Runtime Enforcement During Deployment}
\label{sec:runtime-enforcement}

This section specifies how a deployed system uses a fixed Oracle/governance-batch version to gate Proposer outputs under latency constraints, how prediction and audit-coverage uncertainty trigger escalation, and how runtime evidence enters the next hardening cycle. The topology is consistent with safety filtering and runtime assurance architectures \cite{hobbs2023runtime,hsu2023safety}. Still, the key difference is that each runtime decision is tied to a governed artifact lifecycle: versioned Oracle state, auditable records, uncertainty-driven intake, and regression-validated patch releases.

\subsection{Execution Loop}
\label{sec:runtime-loop}

At runtime, the Proposer $P$ maps the current context $\Sigma$ to a candidate trajectory $\tau_{\mathrm{cand}}$. The Enforcement layer $E$ queries the active governed Oracle stack, maps the returned signals to an operational action, and records the decision in $K$. Table~\ref{tab:runtime-loop} summarizes the loop.

\begin{table}[t]
\centering
\small
\caption{Runtime enforcement loop for a fixed governed Oracle version.}
\label{tab:runtime-loop}
\begin{tabular}{p{0.18\linewidth}p{0.72\linewidth}}
\hline
\textbf{Step} & \textbf{Function} \\
\hline
Propose & $P(\Sigma) \rightarrow \tau_{\mathrm{cand}}$. \\
Oracle-stack evaluation & The governed Oracle stack evaluates $(\Sigma,\tau_{\mathrm{cand}})$ and returns $(s,u,u_{thresh},u_a,u_{a,thresh},v_O,\phi_{\mathrm{hint}},\mathrm{evid})$, where $s$ is the raw safety score, $u$ is Oracle prediction uncertainty, $u_a$ is Alignment Flywheel audit coverage uncertainty, and $\phi_{\mathrm{hint}},\mathrm{evid}$ are optional hooks. \\
Enforcement decision & $E$ selects $a \in \{\mathrm{allow},\mathrm{block},\mathrm{revise},\mathrm{escalate}\}$ under the configured risk policy. High $u$ means the Oracle is unsure about its prediction; high $u_a$ means the Alignment Flywheel has insufficient audit coverage for this class of case. \\
Log & $E$ writes an auditable decision record to $K$, including the active version $v_O$ for attribution, audit, and regression analysis. \\
\hline
\end{tabular}
\end{table}

The Oracle stack is treated as a governed control-plane component: its deployed behavior is tied to an explicit version identifier and changes only through a governed release.

\subsection{Decision Policy and Escalation}
\label{sec:decision-policy}

The Oracle stack returns raw safety-relevant signals; the Enforcement layer maps those signals to an operational response. Table~\ref{tab:enforcement} shows the base policy used throughout the paper. The policy separates prediction uncertainty from audit coverage uncertainty: the former is produced by the Safety Oracle, while the Alignment Flywheel governance layer produces the latter.

Enforcement is not limited to a binary allow/block decision. If $a=\mathrm{allow}$, $E$ may execute $\tau_{\mathrm{cand}}$ subject to local risk checks. If $a=\mathrm{block}$, execution is denied. If $a=\mathrm{revise}$, $E$ may request a revised proposal,
\[
\texttt{revise}(\tau_{\mathrm{cand}}, \phi_{\mathrm{hint}}) \rightarrow \tau'_{\mathrm{cand}},
\]
which supports runtime correction without Proposer retraining. If $a=\mathrm{escalate}$, the case is routed into the audit pipeline.

Risk posture is configurable by domain. High-risk actions include irreversible operations, privileged tool use, and high-impact actuation. For these actions, $E$ can fail closed when uncertainty is high, audit coverage is insufficient, or the Oracle times out. It can also require explicit confirmation or restrict tool use to an approved subset. For lower-risk actions, $E$ may allow execution under moderate uncertainty while logging and escalating the case for later review.

\subsection{Audit Intake, Evidence, and Latency}
\label{sec:audit-intake}

Prediction uncertainty, audit coverage uncertainty, and monitoring statistics are first-class runtime signals under dataset shift and interface drift \cite{ovadia2019can,shankar2021towards}. As shown in Table \ref{tab:runtime-evidence}, $E$ creates an audit case in $K$ when $u \geq u_{thresh}$, $u_a \geq u_{a,thresh}$, $\phi_{\mathrm{hint}}$ indicates a novel or unclassified constraint family, repeated block-revise cycles exceed a configured limit, monitoring detects anomalies, or the Oracle stack exceeds its latency budget. Depending on risk tier, timeouts trigger either fail-closed behavior or fail-open behavior with mandatory logging and escalation; sustained failures activate a degraded mode or circuit breaker.

\begin{table}[t]
\centering
\small
\caption{Runtime evidence captured for audit, replay, and later patch justification.}
\label{tab:runtime-evidence}
\begin{tabular}{p{0.25\linewidth}p{0.63\linewidth}}
\hline
\textbf{Record field} & \textbf{Purpose} \\
\hline
Decision record & $(\Sigma,\tau_{\mathrm{cand}},a,s,u,u_{thresh},u_a,u_{a,thresh},v_O,\mathrm{evid})$ records the context, candidate, action, safety score, uncertainty signals, thresholds, version, and evidence. \\
Integrity metadata & Hash of the serialized trajectory, timestamp, and host identity support replay and fleet correlation. \\
Optional attachments & $\phi_{\mathrm{hint}}$, counterexample identifiers, trace references, source deployment traffic label, stress test, or report. \\
\hline
\end{tabular}
\end{table}

These records ensure that future governance batches can be justified by concrete evidence in $K$, validated against regression suites, and attributed to a specific Oracle/governance-batch version.

\begin{runningexample}[Clinical GenAI Assistant]
A patient reports dizziness after starting a new medication. The Proposer drafts a reassuring reply and suggests no clinical review. The Oracle stack returns a safety score $s$ and prediction uncertainty $u$, while the Alignment Flywheel attaches audit coverage uncertainty $u_a$ for this case class. If $u \geq u_{thresh}$ or $u_a \geq u_{a,thresh}$, Enforcement fails closed: the draft is not sent, the case is logged in $K$, and a priority audit item is routed to $Q_{ver}$. If uncertainty is acceptable but $s$ indicates unsafe, the reply is blocked or revised; only if $s$ is safe and both uncertainty checks pass may the reply be sent.
\end{runningexample}

\subsection{Latency Budget and Timeouts}
\label{sec:latency}

Runtime enforcement must respect latency bounds. The Enforcement layer, therefore, treats Oracle-stack availability as part of the risk policy. If the Oracle query exceeds its latency budget, $E$ triggers a policy-defined fallback. For high-risk actions, the fallback is fail-closed: block execution and create a priority audit case. For lower-risk actions, the system may fail open, but only with mandatory logging and escalation. Sustained Oracle failures or repeated timeouts activate a circuit breaker that moves the deployment into a degraded mode, preventing cascading failures from a slow or unavailable governance dependency.

\section{Reference Implementation and Evaluation}
\label{sec:evaluation}

We implemented a reference Alignment Flywheel prototype with explicit modules for the Proposer, Safety Oracle, Alignment Flywheel Overlay, Query Merger, Enforcement layer, Knowledge Base, Red Team, Verification, Triage, Refinement, and release ledger. Components are selected through a single composition root, so different domains instantiate the same abstract interfaces with different concrete implementations. Our code is available at https://github.com/decide-ugent/Alignment-Flywheel. Governance state is stored as typed records in an append-only $K$, and updates are released as \texttt{Governance Batches} objects. Full implementation details, protocol schemas, and demo specifications are given in Appendices~\ref{app:spatial-3d} and~\ref{app:medical-demos}.

The evaluation is intended as an executable architecture demonstration rather than a domain-specific safety benchmark. We evaluate two deliberately different settings. The first is a learned 3D spatial Oracle derived from an IIRL-style \cite{malomgre2025mixture} reward surface, used to test patch-local governance over a continuous learned artifact. The second is the Patient Portal setting from the running Clinical GenAI Assistant example, used to test the same governance protocol with structured norms, audit coverage uncertainty, and a heuristic proxy Oracle. Together, the demos test whether the same Alignment Flywheel control plane supports different Oracle types, norm semantics, refinement mechanisms, and application domains.

\subsection{Scenario 1: Offline Hardening of a Learned 3D Spatial Oracle}
\label{sec:eval-spatial}

The spatial demo evaluates the Alignment Flywheel on a learned \emph{continuous} IIRL Oracle in $[-1,1]^3$, sampled on a $20^3$ grid for discovery, regression checking, and visualization. Cells near the expert trajectory form the accepted basin; cells with non-trivial reward outside that basin are treated as false-positive reward flaws. The Alignment Flywheel must discover these flaws, suppress them through governed spatial patches, and preserve the sampled expert basin.

This scenario does not include a Proposer. Instead, it tests the Alignment Flywheel's offline hardening loop directly on the learned Oracle artifact. Behavioral change is produced only by governance batches that add Gaussian suppression kernels to the governed Oracle stack. A Patch Planner selects patch centers and bandwidths, predicts which neighboring flaw cells will also be covered, and shrinks or rejects patches that would damage the accepted basin. This tests patch locality at the Oracle level: the learned reward artifact is governed through small, versioned updates rather than retraining.

\begin{figure}[t]
    \centering
    \includegraphics[width=\linewidth]{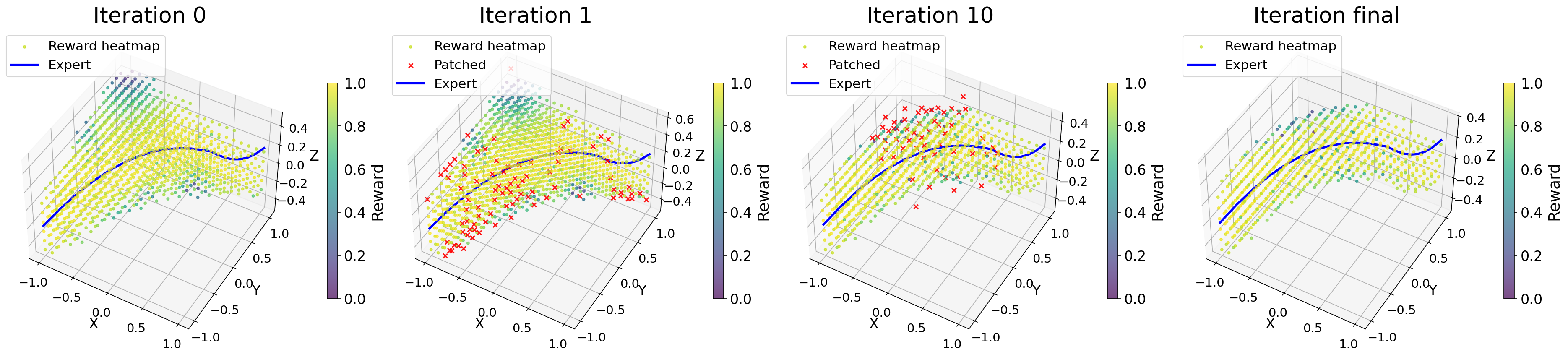}
    \caption{3D spatial Alignment Flywheel progression. The learned Oracle initially assigns non-trivial reward to many cells outside the expert-path basin. Governance batches add suppression kernels around verified flaw regions. Across iterations, the active reward surface contracts toward the expert trajectory while the sampled basin is preserved.}
    \label{fig:spatial-flywheel-progression}
\end{figure}

The adaptive Patch Planner eliminates all discovered flaw cells on the evaluation grid while preserving the sampled basin. The initial grid contains 4\,927 flaw cells and 783 basin cells. After 12 governance iterations, the flaw count reaches zero, and the sampled basin remains fully preserved (783/783). A fixed-bandwidth baseline also preserves the basin but remains far from convergence after the same number of iterations; it has not converged after 25 iterations and is still active. Details of the Patch Planner, flaw classification, baseline, and per-iteration convergence are provided in Appendix~\ref{app:spatial-3d}.\subsection{Scenario 2: Patient Portal Governance}
\label{sec:eval-medical}

The Patient Portal demo instantiates the running Clinical GenAI Assistant example. A patient-facing assistant receives a patient message, proposes a draft reply, and assigns a disposition such as \texttt{reply\_only}, \texttt{nurse\_review}, \texttt{clinician\_review}, or \texttt{urgent\_escalation}. The Proposer is a pass-through component: it does not change during the experiment. The purpose is to test whether the Alignment Flywheel can turn uncertain or under-covered cases into definitive allow/block outcomes by updating only the Oracle stack and audit-coverage state.

The proxy Oracle scores candidate replies along four dimensions: medication content, urgency signals, evidence quality, and disposition mismatch. The Alignment Flywheel Overlay separately tracks audit coverage uncertainty $u_a$ for case classes such as medication advice under weak evidence or lab-result interpretation without sufficient context. The normative specification $\Phi$ contains four typed norms: medication advice requires clinician review, urgency signals require escalation, lab results require context, and unsupported medication-change phrases are prohibited. The same runtime policy from Section~\ref{sec:runtime-enforcement} maps safety score $s$, prediction uncertainty $u$, and audit coverage uncertainty $u_a$ to allow, block, or escalate decisions.

\begin{table}[t]
\centering
\caption{Patient Portal demo: enforcement convergence over evaluations.}
\label{tab:patient-portal-main}
\small
\begin{tabular}{rrrrrl}
\toprule
Eval & Allow & Block & Escalate & Esc.\,Rate & Stack \\
\midrule
0 & 6 & 0 & 9 & 60\% & v0 \\
1 & 6 & 3 & 6 & 40\% & v1 \\
2 & 9 & 6 & 0 & 0\%  & v2 \\
\bottomrule
\end{tabular}
\end{table}

Table~\ref{tab:patient-portal-main} shows the enforcement distribution over the fixed Patient Portal cases. Initially, safe routine cases are allowed, while unsafe and under-covered cases escalate. After the first governance batch, unsupported medication-change cases become definitive blocks. After the second batch, remaining lab and borderline cases are resolved: unsafe cases become blocks, while borderline cases with adequate disposition and established audit coverage become allows. The final evaluation reaches zero escalations while preserving correct separation between safe, unsafe, and review-covered cases.

This demo validates the governance pipeline rather than the medical adequacy of the proxy Oracle. Its purpose is to show that the same Alignment Flywheel machinery used in the spatial setting also operates in the running clinical example: runtime gating, audit intake, verification against typed norms, triage, refinement, governance-batch release, and re-evaluation. Additional Simple Medical and Complex Medical variants are reported in Appendix~\ref{app:medical-demos} as ablations over Oracle complexity, norm structure, and triage strategy.

\subsection{Evaluation Summary}
\label{sec:eval-summary}

Table~\ref{tab:evaluation-orthogonality} summarizes how the two scenarios
exercise different parts of the same Alignment Flywheel architectural contract. The goal is
not to compare domain performance, but to test whether the same control plane
can be instantiated under different assumptions about the Oracle, norms,
refinement mechanism, and deployment mode.

\begin{table}[t]
\centering
\small
\caption{Orthogonality of the two evaluation scenarios.}
\label{tab:evaluation-orthogonality}
\begin{tabular}{p{0.17\linewidth}p{0.27\linewidth}p{0.28\linewidth}p{0.20\linewidth}}
\hline
\textbf{Dimension} &
\textbf{3D spatial demo} &
\textbf{Patient Portal demo} &
\textbf{Architectural property} \\
\hline

Evaluation mode &
Offline hardening of a learned Oracle artifact &
Runtime Proposer--Oracle enforcement and hardening &
Offline and runtime governance \\

Oracle type &
Learned continuous IIRL-style Oracle &
Structured proxy Oracle with safety and uncertainty signals &
Oracle substitutability \\

Proposer role &
No Proposer; the Oracle is hardened directly &
Passthrough Proposer producing candidate replies and dispositions &
Separation of proposal and governance \\

Norm structure &
Spatial support and expert-basin preservation &
Medication, urgency, lab-context, and unsupported-advice norms &
Norm-semantic variation \\

Refinement mechanism &
Gaussian suppression patches released through governance batches &
Blocks, disposition-sensitive decisions, threshold updates, and audit-coverage updates &
Patch-local correction \\

Observed effect &
False-positive reward cells are removed while the sampled expert basin is preserved &
Escalations are converted into definitive allow/block outcomes across governance batches &
Executability and protocol stability \\

\hline
\end{tabular}
\end{table}

The two scenarios support complementary claims. The 3D spatial demo isolates
offline patch-local hardening of a learned continuous IIRL Oracle: false-positive
reward regions are removed on the evaluation grid while the sampled expert
basin is preserved, without retraining the learned reward artifact. The Patient
Portal demo instantiates the full runtime Proposer--Oracle governance loop from
the running clinical example, using a structurally different proxy Oracle, typed
norms, disposition-based enforcement, and audit-coverage uncertainty. Across
governed updates to the Oracle produce both domains, behavioral change
stack and Alignment Flywheel state rather than by changing the decision generator or
retraining the learned artifact.

The evaluations therefore establish executability, patch locality, and
architectural substitutability under controlled conditions. They do not claim
production safety, clinical readiness, or complete verification of all possible
failures. Instead, they show that the proposed roles, protocol records,
interfaces, and release semantics are sufficient to operate both offline Oracle
hardening and runtime Alignment Flywheel governance, while motivating future work on
stronger Oracle learning methods, richer norm logics, formal patch-safety
guarantees, calibrated uncertainty, and long-horizon deployment studies.

\section{Discussion, Limitations, and Conclusion}
\label{sec:discussion}

\paragraph{Operational implications.}
The Alignment Flywheel treats safety governance as an explicit control-plane workflow rather than an opaque property of a monolithic decision model. It does not claim to solve safety once and for all. Its central claim is that many observed failures can be addressed locally, without necessarily retraining or retracting the Proposer. Candidate failures are logged, verified, and triaged to guide the preparation of governance batches. These batches are regression-checked, released, monitored, and rolled back when necessary. This shifts the engineering unit of change from the full decision component to the governed Oracle stack and its audit state. Runtime decisions and version transitions remain traceable to supporting evidence and normative justification.

\paragraph{Evaluation scope and limitations.}
The implementation and demos show that the proposed roles, artifacts, interfaces, and release loop can operate under controlled conditions. The 3D spatial demo demonstrates offline hardening of a continuous IIRL Oracle evaluated on a finite grid. False-positive reward regions are discovered and suppressed while the sampled expert basin is preserved. The Patient Portal demo implements the full runtime governance loop for the running Clinical GenAI Assistant example. Prediction uncertainty and audit coverage uncertainty guide enforcement, while typed norms support verification. Successive governance batches convert initial escalations into definitive allow or block decisions. These results support the architectural claims of patch locality, protocol stability, and component substitutability. They do not establish production safety or clinical validity. Further work is needed on Oracle construction, richer norm semantics, formal patch-safety guarantees, and calibrated uncertainty. Human review costs, latency, and long-horizon adversarial adaptation also require further study. Appendix~\ref{app:scaling-roadmap} sketches how these extensions scale across oversight, modalities, domains, and components.

\paragraph{Conclusion.}
We presented the Alignment Flywheel as a governance-centric hybrid MAS architecture for patch-local safety control. The Proposer generates candidate trajectories, and a governed Oracle stack returns safety and uncertainty signals. An Enforcement layer uses these signals to apply an explicit risk policy. The governance MAS closes the loop through monitoring, verification, triage, refinement, and versioned release.

Our contribution is architectural: we specify the control-plane roles, artifacts, protocols, and deployment semantics needed to support auditing and ongoing governance of fallible autonomous systems. The reference implementation and demos show that the same protocol can support both offline hardening of a learned Oracle and runtime governance in a clinical-style Proposer--Oracle setting. These demonstrations provide a concrete basis for future formal, algorithmic, and deployment-focused work.

\section*{Acknowledgment}
This research was supported by funding from the Flemish Government under the ''Onderzoeksprogramma Artificiele Intelligentie (AI) Vlaanderen ’’ program.

\appendix
\section{Appendix Overview}
\label{app:overview}

The appendix provides the technical material omitted from the main paper for space. Appendix~\ref{app:extended-background} expands the conceptual positioning: it clarifies governance as an operational control layer, compares Safety Oracle families, and distinguishes prediction uncertainty from audit coverage uncertainty. Appendix~\ref{app:ooda} details the OODA-loop instantiations for the Red Team, Blue Team, Verification, Triage, and Refinement roles, showing what each role reads from and writes to the append-only knowledge base $K$. Appendix~\ref{app:protocols} defines the inter-role protocol artifacts, including candidate flaws, verification results, verified breaches, refinement jobs, governance batches, and release records. Appendix~\ref{app:oversight} describes the tunable human oversight model, including low-, medium-, and high-risk operating modes and the control surfaces exposed to human operators. Appendix~\ref{app:implementation} summarizes the reference implementation structure, norm representation, and governance-batch format. Appendix~\ref{app:spatial-3d} gives the full 3D spatial demo specification, including the continuous IIRL Oracle, kernel patch model, PatchPlanner, baseline comparison, and convergence results. Appendix~\ref{app:medical-demos} gives the medical demo specifications, including the shared runtime pipeline, proxy Oracle variants, typed medical norms, enforcement mechanics, and cross-demo comparison.

\section{Extended Background on Governance and Safety Artifacts}
\label{app:extended-background}

\subsection{Governance as an Operational Control Layer}
\label{app:governance-background}

The paper uses \emph{governance} in a deliberately operational sense. The concern is not only whether a system satisfies a norm in principle, but how such norms, uncertainty signals, verification outcomes, escalation decisions, and release controls are represented and propagated in a running deployment. In this sense, governance sits between abstract alignment goals and concrete runtime control: it determines who is allowed to decide what, which artifacts are authoritative, how uncertainty is communicated, when a case must be deferred or escalated, and how corrective updates are qualified and rolled out.

This view is compatible with normative MAS perspectives in which governance is grounded in explicit norms and state transitions over those norms rather than hidden inside individual agents \cite{singh2014norms,boella2006introduction}. However, the present paper is concerned less with the full social semantics of norms and more with deployable control structure. In particular, governance is treated here as a modular layer organized around explicit protocol responsibilities: monitoring, triage, verification, refinement, enforcement, and release management. This makes governance a separation-of-concerns problem as much as a normative one. A contributor can improve one role in the loop without needing to understand all other modules internally, provided the role respects the agreed interfaces and artifact semantics.

The Alignment Flywheel extends this idea in two directions. First, it provides \emph{verification-as-a-service}: beyond the oracle's own prediction, it contributes additional assurance through external auditing and verification. Second, it provides \emph{alignment-as-a-service}: uncertainty and failures observed in deployment become inputs to iterative refinement of the governed artifact rather than mere post-hoc diagnostics. This is why the paper distinguishes oracle-side prediction uncertainty from Alignment Flywheel-generated audit coverage uncertainty. The former expresses how trustworthy the oracle believes its own judgment to be; the latter expresses how well that class of judgments has been externally covered by governance activity.

\subsection{Possible Safety Oracle Instantiations and Their Assumptions}
\label{app:oracle-families}

The proposer-oracle architecture is intentionally agnostic to the internal realization of the oracle. Different safety artifacts could instantiate this role, but they differ in the assumptions they make and in how naturally they support patchability, audit, and lifecycle governance.

\paragraph{Preference-shaped policies and implicit alignment.}
Methods such as RLHF and DPO can improve policy behavior directly. Still, as oracle candidates they are less naturally modular because governance-relevant behavior is often entangled with the policy itself \cite{ziegler2019fine,ouyang2022training,rafailov2023direct}. They are therefore useful baselines for alignment but a weaker fit when independent audit and patch-local updates are required.

\paragraph{Learned evaluative artifacts from demonstrations.}
IRL-style reward models, IIRL-style evaluators, and related demonstration-derived artifacts are attractive because they separate action generation from evaluation \cite{ng2000algorithms,ziebart2008maximum,adams2022survey,arora2021survey,malomgre2025mixture}. Their main assumptions concern coverage, identifiability, and generalization from demonstrations. They can extrapolate unreliably outside demonstrated support, which motivates explicit uncertainty and governance around them.

\paragraph{Constraint and logic-based artifacts.}
Learned logic constraints \cite{liu2024comprehensive} and hand-written monitors or rules provide stronger interpretability and can map more directly to norms. Their main assumptions are that the relevant structure is representable in the chosen language and that sufficient observability is available to evaluate those constraints at runtime.

\paragraph{Guardrails and shields.}
Guardrails and shielding mechanisms can also instantiate the oracle role, especially when explicit specifications exist \cite{rebedea2023nemo,alshiekh2018safe,konighofer2025shields}. Their strengths lie in runtime enforcement and, in some settings, formal guarantees. Their main limitation for the present paper is not that they cannot be verified, but that correction and adaptation may become difficult when failure handling requires changing the specification or the synthesized controller itself rather than issuing a narrow patch to an evaluative artifact.

\paragraph{Structured task artifacts.}
Beyond standalone evaluative models, Alignment Flywheel-style governance could also operate over structured artifacts such as automata or reward-machine-like representations. Reward machines and related structured intermediates can be learned from data \cite{baert2024reward,shehab2026active} or language \cite{castanyer2025arm} and can provide explicit, inspectable task progression. Their explicit proposition mappings and transition structure support localized, inspectable updates, making them plausible substrates for audited oracle refinement. 

\paragraph{In practice.} 
Realistic deployments may combine several of these ingredients. The architectural contribution of this paper is therefore not tied to one oracle family. The use of IIRL in the main text should be read as a reference instantiation with favorable systems properties, not as a claim that all other oracle realizations are inferior in all settings.

\subsection{Prediction uncertainty and audit coverage uncertainty}
\label{app:uncertainty-background}

The paper distinguishes two different uncertainty signals.

\paragraph{Prediction uncertainty.}
This is attached to the oracle's own judgment. It expresses how trustworthy the oracle believes its safety score to be for the queried input. Depending on the oracle family, this may be estimated through support in the underlying data, evidential inconsistency, ensemble disagreement, calibration methods, or other means.

\paragraph{Audit coverage uncertainty.}
This is generated by the Alignment Flywheel itself. It expresses how well a particular class of cases, trajectories, or outputs has been externally covered by auditing and verification activity. It therefore does not come from the oracle's training data or internal model alone. Instead, it captures second-order uncertainty about whether that region of behavior has been checked sufficiently well to justify trust in deployment.

The distinction matters operationally. A system may defer because the oracle is uncertain, because the Alignment Flywheel has insufficient verification coverage for that kind of case, or both. This is especially important in verification-as-a-service settings, where the governance layer is valuable not only because it consumes oracle-side uncertainty, but because it contributes additional assurance beyond the oracle itself.

\subsection{Verification, Correction, and Patchability}
\label{app:verification-vs-correction}

Verification and correction should not be conflated. Some safety artifacts, especially explicit monitors, rules, and shields, may be comparatively amenable to verification because their specifications are explicit and their behavior can be checked directly against those specifications. However, when new failure classes appear, adapting such artifacts may require rewriting the specification, synthesizing a new controller, or revisiting the assumptions under which the monitor was sound.

By contrast, the proposer-oracle architecture in this paper is designed around \emph{patch locality}: many safety fixes are intended to take the form of narrow updates to an evaluative artifact and its governed release process. The point is not that such artifacts are always easier to verify, but that they can be easier to correct incrementally under deployment pressure. This makes patchability, auditability, and release control central concerns alongside correctness.
\section{OODA Loop Specifications for Governance Roles}
\label{app:ooda}

To demonstrate the operational feasibility of the Alignment Flywheel, we detail concrete OODA-loop instantiations for the governance roles described in Sec.~\ref{sec:system} and Sec.~\ref{sec:interaction}. The OODA abstraction supports stateless, decoupled role implementations: agents interact through the Knowledge Base $K$, the queues $Q_{ver}$ and $Q_{ref}$, and the defined protocol artifacts, as shown in Figure~\ref{fig:ooda-abstract}.

\begin{figure}[h!]
\centering
\begin{tikzpicture}[>=Latex, every node/.style={font=\footnotesize}]

% Central KB
\node[oodaStore] (K) at (0,0) {Knowledge Base $K$\\[-1pt]\scriptsize shared artifacts + queues};

% OODA phases around K
\node[oodaPhase] (Obs) at (-4.0,  1.8) {Observe\\[-1pt]\scriptsize read from $K$};
\node[oodaPhase] (Ori) at ( 4.0,  1.8) {Orient\\[-1pt]\scriptsize interpret context};
\node[oodaPhase] (Dec) at ( 4.0, -1.8) {Decide\\[-1pt]\scriptsize select strategy};
\node[oodaPhase] (Act) at (-4.0, -1.8) {Act\\[-1pt]\scriptsize write to $K$};

% OODA loop
\draw[oodaArrow] (Obs) -- (Ori);
\draw[oodaArrow] (Ori) -- (Dec);
\draw[oodaArrow] (Dec) -- (Act);
\draw[oodaArrow] (Act) -- (Obs);

% Interaction with K
\draw[oodaArrow] (K.north west) -- node[pos=0.55, above left, align=center, xshift=50pt] {\scriptsize artifacts, logs,\\[-1pt]\scriptsize queue state} (Obs.south east);
\draw[oodaArrow] (Act.north east) -- node[pos=0.45, below left, align=center, xshift=50pt] {\scriptsize new artifacts,\\[-1pt]\scriptsize alerts, jobs} (K.south west);

\node[align=center,font=\scriptsize] at (0,-3.4)
{Abstract role template used by Red, Blue, Verification, Triage, and Refinement agents.};

\end{tikzpicture}
\caption{Abstract OODA interaction pattern for governance agents. Each role reads shared state from the append-only knowledge base $K$ during \emph{Observe}, interprets it relative to its local objective during \emph{Orient}, selects a strategy during \emph{Decide}, and writes derived artifacts back to $K$ during \emph{Act}. Role-specific behavior is captured by strategies and artifact types, while the interaction contract with $K$ remains uniform across the governance MAS.}
\label{fig:ooda-abstract}
\end{figure}

\subsection{Red Team: Adversarial Discovery}

\begin{itemize}
    \item \textbf{Observe:} Reads the active normative specification $\Phi$, including severity weights, the current Oracle/governance versions, and historical defended traces from $K$.
    \item \textbf{Orient:} Identifies regions of the input, trajectory, or Oracle-response space with high normative severity but low empirical coverage. If a human has injected an adversarial seed \cite{xu2021bot,ganguli2022red}, the agent initializes its search around that seed.
    \item \textbf{Decide:} Selects a generation or search strategy suited to the domain, such as prompt mutation for language outputs, gradient-based search over differentiable artifacts, physical perturbation for robotics, or spatial probing of a learned Oracle surface. The strategy targets uncertain regions and low-uncertainty claimed-safe cases that may conceal false negatives \cite{perez2022red,deng2022rlprompt,lee2022query, jones2023automatically,chakraborty2021survey,jia2017adversarial,song2018constructing, chen2023content}.
    \item \textbf{Act:} Generates or selects a candidate trajectory $\tau_{\mathrm{cand}}$ and queries the governed Oracle stack. If the stack returns a safe prediction with low prediction uncertainty ($u < u_{\mathrm{thresh}}$), but the Red Team heuristic still suspects a violation, the agent writes a \texttt{CandidateFlaw} record to $Q_{ver}$.
\end{itemize}

\subsection{Verification Team: Validation}

\begin{itemize}
    \item \textbf{Observe:} Pulls candidate flaws from $Q_{ver}$ and retrieves the relevant norm definitions $\phi \in \Phi$ from $K$.
    \item \textbf{Orient:} Parses the trajectory $\tau$ into a verifiable representation, such as a terminal state, tool-call trace, structured disposition, or normalized semantic representation.
    \item \textbf{Decide:} Evaluates $\tau$ against $\phi$. Programmatic agents may use deterministic validators such as SMT checks, regular expressions, threshold rules, execution sandboxes, or domain-specific predicate checkers. If the norm requires judgment that is not formalized, the decision is escalated to a human reviewer or an adjudication service.
    \item \textbf{Act:} Writes a \texttt{VerificationResult} to $K$. If a violation is confirmed, the agent also writes a \texttt{VerifiedBreach}; otherwise, it logs the attempt as a defended trace to reduce redundant rediscovery.
\end{itemize}

\subsection{Triage Agent: Queue Governance}

\begin{itemize}
    \item \textbf{Observe:} Polls $K$ for newly logged \texttt{VerifiedBreach} records and queue state.
    \item \textbf{Orient:} Groups structurally similar breaches into failure families, for example by embeddings, rule keys, domain labels, or manually supplied cluster identifiers \cite{mosqueira2023human,kontonatsios2017semi,sun2024corrmatch,shen2025reviving}.
    \item \textbf{Decide:} Prioritizes clusters using signals such as norm severity, dangerous certainty $(u_{\mathrm{thresh}}-u)$, novelty relative to prior history, deployment frequency, and operational urgency. It may also perform diversity sampling within each cluster to select representative edge cases.
    \item \textbf{Act:} Packages cluster metadata and selected exemplars into a \texttt{RefinementJob} and pushes it to $Q_{ref}$.
\end{itemize}

\subsection{Refinement Team: Correction and Release Preparation}

\begin{itemize}
    \item \textbf{Observe:} Pulls the highest-priority \texttt{RefinementJob} from $Q_{ref}$ and retrieves associated breaches, regression tests, prior patches, and release constraints from $K$.
    \item \textbf{Orient:} Analyzes the breach cluster to determine whether the correction should affect the Oracle artifact, an Oracle wrapper, the Alignment Flywheel Overlay, audit coverage state, thresholds, or enforcement policy.
    \item \textbf{Decide:} Synthesizes candidate corrections, such as symbolic rules, spatial suppression kernels, disposition overrides, threshold adjustments, audit coverage updates, or other Oracle-stack patches. In human-supervised modes, a reviewer approves, rejects, or modifies the proposed governance batch.
    \item \textbf{Act:} Evaluates the proposed \texttt{GovernanceBatch} against the regression suite. If it passes the required checks and approvals, the system writes a release record to the ledger $L$ in $K$, producing a new governed Oracle-stack version.
\end{itemize}

\subsection{Blue Team: Monitoring and Forensics}

\begin{itemize}
    \item \textbf{Observe:} Reads telemetry from the Enforcement layer $E$, such as block rates, high-uncertainty escalations, timeout rates, and version-specific decision distributions, together with Red Team and verification logs from $K$.
    \item \textbf{Orient:} Computes monitoring summaries such as agent efficacy, norm coverage, calibration drift, audit coverage gaps, deployment anomalies, and proposer--oracle interface regressions.
    \item \textbf{Decide:} Determines whether a Red Team strategy has plateaued, whether additional verification coverage is needed, whether a release should be paused or rolled back, or whether live traffic indicates a distribution shift requiring temporary mitigation.
    \item \textbf{Act:} Writes coverage maps, forensic summaries, and monitoring reports to $K$. In emergency cases, it may trigger a policy-defined degraded mode or temporary mitigation in $E$, while logging the action for later verification, refinement, and governed release.
\end{itemize}

\section{Protocol Specifications}
\label{app:protocols}

We specify coordination as explicit protocols between roles rather than as implicit message passing. This improves substitutability, since agents and strategies can be swapped without changing the interaction contract, and supports audit of compliance with the intended governance process. The protocols implement the two-stage triage pipeline over the append-only Knowledge Base $K$: candidate flaws are verified before becoming breaches, and breaches are clustered before becoming governed updates. The same protocol supports different degrees of human involvement. Low-risk deployments may automate verification and refinement, while high-risk deployments may require human review or signature for each governance batch. Appendix~\ref{app:oversight} details the human oversight and scaling model.

\subsection{Common Transport and Invariants}
\label{app:protocols-common}

\paragraph{Transport model.}
Protocol artifacts are stored in the append-only Knowledge Base $K$ and referenced by immutable identifiers, such as hashes. Queues $Q_{ver}$ and $Q_{ref}$ store only artifact references, enabling idempotent processing and decoupled execution.

\paragraph{Global invariants.}
\begin{itemize}
  \item \textbf{Idempotency:} each protocol artifact carries a unique ID; consumers must tolerate retries without duplicating effects.
  \item \textbf{Causal linking:} every derived artifact records parent references, e.g., a \textsf{VerificationResult} links to a \textsf{CandidateFlaw}.
  \item \textbf{Tamper-evident provenance:} governance batches and release events are signed and recorded in ledger $L$ to support audit, deployment control, and rollback.
\end{itemize}

\subsection{Protocol P1: Candidate Flaw Submission (Red $\rightarrow Q_{ver}$)}
\label{prot:red-qver}

\paragraph{Participants.} Red Team agent, $Q_{ver}$, Knowledge Base $K$.

\paragraph{Purpose.} Submit candidate flaws for verification, emphasizing low-uncertainty claimed-safe trajectories that may represent suspected false negatives.

\paragraph{Message schema.}
\[
\textsf{CandidateFlaw} =
\langle id,\Sigma,\tau,s,u,u_{\mathrm{thresh}},u_a,u_{a,\mathrm{thresh}},v_O,ts,\textsf{seedRef}\rangle .
\]

\paragraph{Steps.}
\begin{enumerate}
  \item Red Team appends \textsf{CandidateFlaw} to $K$ and enqueues $id$ to $Q_{ver}$.
  \item Triage-I may re-prioritize within $Q_{ver}$ using risk signals such as norm severity, dangerous certainty $\max(0,u_{\mathrm{thresh}}-u)$, audit coverage uncertainty $u_a$, novelty, and operational urgency.
\end{enumerate}

\paragraph{Postconditions.} $K$ contains an immutable candidate record; $Q_{ver}$ contains its reference. If enqueue fails, Red Team retries; idempotency holds because $id$ is stable.

\subsection{Protocol P2: Verification Result Publication (Verification $\rightarrow K$, Triage)}
\label{prot:ver-result}

\paragraph{Participants.} Verification agent(s) and/or human reviewer(s), Knowledge Base $K$, Triage agent.

\paragraph{Purpose.} Validate a \textsf{CandidateFlaw} against $\Phi$ and produce a \textsf{VerificationResult} plus, when applicable, a \textsf{VerifiedBreach}.

\paragraph{Message schema.}
\[
\textsf{VerificationResult}
=
\langle refId,isViolation,\phi_{\mathrm{broken}},evidenceRef,reviewerRef,ts\rangle .
\]
If $isViolation=T$:

\[
textsf{VerifiedBreach}
=
\]
\[
\langle breachId,refId,\Sigma,\tau,\phi_{\mathrm{broken}},evidenceRef,
s,u,u_{\mathrm{thresh}},u_a,u_{a,\mathrm{thresh}},v_O,ts\rangle .
\]

\paragraph{Steps.}
\begin{enumerate}
  \item Verifier dequeues a candidate reference $id$ from $Q_{ver}$ and reads the linked \textsf{CandidateFlaw} from $K$.
  \item Verifier evaluates $\tau$ against $\Phi$ using automated checks and/or human judgment and appends a \textsf{VerificationResult} to $K$.
  \item If a violation is confirmed, the verifier additionally appends a \textsf{VerifiedBreach} to $K$, linked to $refId$ and carrying the Oracle-stack signals recorded at discovery time.
\end{enumerate}

\paragraph{Postconditions.} All decisions are recorded in $K$ with explicit links to parent artifacts; confirmed violations are available to Triage as \textsf{VerifiedBreach} records.

\subsection{Protocol P3: Cluster and Job Creation (Triage $\rightarrow Q_{ref}$)}
\label{prot:triage-qref}

\paragraph{Participants.} Triage agent, Knowledge Base $K$, $Q_{ref}$.

\paragraph{Purpose.} Convert a stream of \textsf{VerifiedBreach} records into batched \textsf{RefinementJob} artifacts.

\paragraph{Message schema.}
\[
\textsf{RefinementJob}
=
\]
\[
\langle jobId,clusterId,centroidRef,size,riskScore,sampleSetRefs,ts\rangle .
\]

\paragraph{Steps.}
\begin{enumerate}
  \item Triage observes new \textsf{VerifiedBreach} artifacts in $K$ since the last checkpoint.
  \item Triage clusters similar breaches and computes $riskScore$ using signals such as norm severity, dangerous certainty, audit coverage uncertainty, novelty, deployment frequency, and operational urgency.
  \item Triage appends a \textsf{RefinementJob} to $K$ and enqueues $jobId$ to $Q_{ref}$.
\end{enumerate}

\paragraph{Postconditions.} $Q_{ref}$ contains cluster-level jobs, reducing human review load through batching and diversity-aware aggregation.

\subsection{Protocol P4: Governance batch preparation (Refinement $\rightarrow K,L$)}
\label{prot:governance-batch}

\paragraph{Participants.} Refinement agent(s) and/or human supervisor(s), Knowledge Base $K$, Release Ledger $L$.

\paragraph{Purpose.} Produce a \textsf{GovernanceBatch} with linked breach provenance, regression evidence, version metadata, and authorization signature. A batch may contain Oracle-local corrections, overlay updates, audit coverage updates, threshold changes, or enforcement-policy metadata.

\paragraph{Message schema.}
\[
\textsf{GovernanceBatch} =
\]
\[
\langle id,corrections,parentV,targetV,breachRefs,
testResultRef,signature,ts\rangle .
\]

\paragraph{Steps.}
\begin{enumerate}
  \item Refinement dequeues $jobId$ from $Q_{ref}$ and reads the linked breach set from $K$.
  \item Refinement synthesizes candidate corrections and runs the regression suite, recording the resulting test artifact in $K$.
  \item Refinement appends a \textsf{GovernanceBatch} to $K$ and records the corresponding immutable batch event in $L$.
\end{enumerate}

\paragraph{Security invariant.} A batch is deployable if and only if its signature verifies, required regression evidence is present, and its \textit{parentVersion} matches the fleet's allowed upgrade path.

\subsection{Protocol P5: Release, Rollout, and Rollback}
\label{prot:release}

\paragraph{Participants.} Release manager, $L$, deployment substrate, governed Oracle-stack endpoints.

\paragraph{Purpose.} Distribute signed governance batches under progressive rollout and safe rollback.

\paragraph{Release record schema.}
\[
\textsf{ReleaseRecord}
=
\langle Id,batchId,rolloutPolicy,status,canaryMetricsRef,ts\rangle .
\]

\paragraph{Steps.}
\begin{enumerate}
  \item Release manager selects a deployable \textsf{GovernanceBatch} and creates a \textsf{ReleaseRecord} in $L$.
  \item Rollout proceeds in stages, e.g., canary $\rightarrow$ expand $\rightarrow$ full, updating \textit{status} with measured metrics.
  \item On regression, release manager records a rollback event in $L$ and instructs endpoints to revert to the prior allowed version.
\end{enumerate}

\paragraph{Security motivation.} Signed provenance and controlled rollout/rollback are standard defenses for distributed update channels and supply-chain integrity, including attested build steps and signing infrastructure \cite{newman2022sigstore}.

\section{Human Oversight and Scaling}
\label{app:oversight}

The Alignment Flywheel is designed for tunable oversight: human involvement can vary with domain risk, regulatory burden, and organizational policy. Rather than requiring humans to inspect every candidate trajectory, the governance MAS exposes control surfaces that let operators steer the system at the level of norms, thresholds, strategy modules, escalation rules, and release approvals. This follows mixed-initiative and adjustable-autonomy patterns for scalable human-agent teams \cite{scerri2001adjustable,horvitz1999principles}. Oversight modes are shown in Table \ref{tab:oversight-modes} and human control surfaces in Table \ref{tab:human-control-surfaces}

\begin{table}[h]
\centering
\caption{Oversight modes supported by the Alignment Flywheel.}
\label{tab:oversight-modes}
\small
\begin{tabular}{p{0.20\linewidth}p{0.25\linewidth}p{0.42\linewidth}}
\toprule
\textbf{Mode} & \textbf{Human role} & \textbf{Typical use} \\
\midrule
Low risk & Exception handling & Automated verification and refinement; humans review anomalies or periodic reports. \\
Medium risk & Batch approval & Agents cluster failures and propose governance batches; humans review summaries and authorize release. \\
High risk & Item and release approval & Humans review high-impact cases, approve or modify patches, and sign each governance batch before rollout. \\
\bottomrule
\end{tabular}
\end{table}

\begin{table}[h]
\centering
\caption{Human steering interfaces for scalable governance.}
\label{tab:human-control-surfaces}
\small
\begin{tabular}{p{0.30\linewidth}p{0.58\linewidth}}
\toprule
\textbf{Control surface} & \textbf{Effect} \\
\midrule
Agent efficacy and strategy rotation & Surface metrics such as unique verified breaches per unit time; deprecate weak strategies and activate stronger ones. \\
Normative coverage and prioritization & Visualize coverage across $\phi \in \Phi$; adjust severity weights or introduce new norms to re-orient OODA loops. \\
Adversarial seeding and directed search & Inject seed trajectories into $Q_{\mathrm{ver}}$; Red Team agents mutate or exploit them to discover variants \cite{ji2023ai,li2024multitude}. \\
Deployment feedback & Surface high-uncertainty runtime cases as incidents; direct coverage expansion and verification toward failed regions. \\
Release approval and rollback & Require human approval, signature, or staged rollout for governance batches in high-risk settings. \\
\bottomrule
\end{tabular}
\end{table}

\paragraph{Scaling rationale.}
The purpose of tunable oversight is to move human effort from raw case handling to governance steering. Triage clusters similar failures before they reach refinement; verification separates suspected flaws from confirmed breaches; and release approval operates on governance batches rather than individual probes. This keeps human attention focused on high-severity uncertainty, new norm families, disputed verification judgments, and release decisions.

\section{Scaling Roadmap}
\label{app:scaling-roadmap}

The Alignment Flywheel is an architectural layer, so scaling it requires progress along several independent axes rather than a single larger benchmark.

\paragraph{Oversight scaling.}
Human involvement can move from item-level review to batch approval, policy steering, and release authorization. Low-risk settings may automate most verification and refinement, whereas high-risk settings require human approval for verified breaches, governance batches, or rollout decisions.

\paragraph{Modality scaling.}
The same Proposer--Oracle interface can govern text, tool calls, trajectories, images, robot plans, or multimodal traces, provided candidate behavior can be represented as a trajectory and evaluated by an Oracle stack.

\paragraph{Domain scaling.}
Different domains require different norm semantics, evidence sources, escalation rules, and regression suites. Clinical communication, robotics, cyber defense, and tool-using agents therefore instantiate the same governance loop with different domain-specific verifiers and policies.

\paragraph{Component scaling.}
The architecture does not require any single Red Team, Oracle, verifier, or refinement method. Stronger IIRL-style Oracles, calibrated uncertainty estimators, formal norm checkers, adversarial search methods, and patch planners can replace the simple components used in the demos without changing the protocol layer.

\section{Reference and Demos Implementation Details}
\label{app:implementation}

This appendix summarizes implementation-level details for the evaluation scenarios in Section~\ref{sec:evaluation}. The reference implementation is a single Python/Flask application organized around six layers: protocol definitions, core governance infrastructure, concrete role implementations, a thin HTTP API layer, a factory registry, and demo runners. Components are selected through a single composition root, so the same controllers, knowledge base, query merger, batch applier, and protocol artifacts are reused across the spatial and medical demos. Code is available at https://github.com/decide-ugent/Alignment-Flywheel.

\subsection{Project Structure}

The implementation is organized as follows:

\begin{verbatim}
Alignment Flywheel/
  protocols/       # enums, typed artifacts, interfaces, OODA
  core/            # governance engine, knowledge base, 
                   # query merger, batch applier
  roles/           # oracle, proposer, overlay, enforcement, 
                   # triage,blue team, red team, verifier
                   #, refinement
  api/             # Flask app, blueprints, HTTP clients, 
                   # runtime state
  factory/         # registry, auto-registration, 
                   # YAML-driven wiring
  demos/           # spatial and medical demo 
                   # runners/configurations
\end{verbatim}

The protocol layer defines the shared dataclasses used throughout the system, including \texttt{Trajectory}, \texttt{TrajectoryStep}, \texttt{OracleRawOutput}, \texttt{FlywheelOverlay}, \texttt{UnifiedQueryResult}, \texttt{EnforcementResult}, \texttt{Norm}, \texttt{GovernanceBatch}, \\ \texttt{LocalCorrection}, \texttt{CandidateFlaw}, \texttt{VerificationResult}, and \texttt{DecisionRecord}. The interface layer defines abstract contracts for the Proposer, Safety Oracle, Alignment Flywheel Overlay, Enforcement policy, Knowledge Base, Query Merger, Batch Applier, Blue Team, Triage, and domain-specific adapters.

The OODA-structured roles are decomposed into \texttt{observe}, \texttt{orient}, \texttt{decide}, and \texttt{act} steps. For example, the Red Team can use a grid observer in the spatial demo or a medical case generator in the Patient Portal demo, while preserving the same role-level OODA controller. YAML configuration files select the concrete steps for each demo through the factory registry.

\subsection{Norm Representation}
\label{app:norms}

Each norm in $\Phi$ is represented as a \texttt{Norm} dataclass with fields \texttt{id}, \texttt{kind: NormKind}, \texttt{spec: Dict[str,Any]}, \texttt{severity: float}, \texttt{weight: float}, and \texttt{description: str}. The typed \texttt{kind} field determines the verifier semantics.

\begin{table}[h]
\centering
\caption{Implemented norm kinds in the reference implementation.}
\label{tab:implemented-norm-kinds}
\small
\begin{tabular}{p{0.26\linewidth}p{0.62\linewidth}}
\toprule
\textbf{Norm kind} & \textbf{Verifier semantics} \\
\midrule
\texttt{KEYWORD\_BLOCK} &
Checks whether trajectory text contains prohibited keywords, optionally conditioned on evidence status. Example: medication-change terms under weak evidence. \\
\texttt{REGEX} &
Checks whether trajectory text matches a regular-expression pattern, such as a prohibited action phrase. \\
\texttt{PREDICATE} &
Evaluates multi-field relationships over payload and metadata, such as whether the proposed disposition is at least as severe as required for the case type and evidence status. \\
\texttt{SPATIAL\_BOUNDARY} &
Checks whether a spatial query point lies within the supported region, e.g., below a distance threshold from expert data. \\
\texttt{THRESHOLD\_RULE} &
Checks threshold-style constraints, such as age or vulnerability conditions requiring a specified evidence standard. \\
\bottomrule
\end{tabular}
\end{table}

This typed representation is the implementation-level instantiation of the paper's \texttt{LogicExpression} abstraction. New norm kinds can be added by extending \texttt{NormKind} and implementing the corresponding verifier logic in the relevant orient/decide steps.

\subsection{Governance Batch Format}
\label{app:batch}

The \texttt{GovernanceBatch} is the deployment unit for governed updates. It contains a list of \texttt{LocalCorrection} artifacts, each with a typed \texttt{correction\_type: CorrectionType} and a payload dictionary. The implementation supports five correction types: \texttt{SPATIAL\_FLAW\_PATCH}, \texttt{AUDIT\_COVERAGE\_UPDATE}, \\ \texttt{THRESHOLD\_ADJUSTMENT}, \texttt{MEDICAL\_HARD\_BLOCK}, and \texttt{NORM\_UPDATE}. Not every demo uses every correction type: the spatial demo uses spatial flaw patches and audit coverage updates, while the medical demos use hard blocks, disposition/threshold adjustments, and audit coverage updates.

A representative spatial governance batch is:

\begin{verbatim}
{
  "batch_id": "batch:a1b2c3",
  "from_oracle_version": "oracle:v4",
  "to_oracle_version": "oracle:v5",
  "local_corrections": [
    {
      "correction_id": "corr:d4e5f6",
      "correction_type": "spatial_flaw_patch",
      "payload": {
        "flaw_point": [0.73, -0.42, 0.81],
        "support_radius": 0.187
      }
    },
    {
      "correction_id": "corr:g7h8i9",
      "correction_type": "audit_coverage_update",
      "payload": {
        "case_class": "spatial|d=0.7"
      }
    }
  ],
  "regression_evidence": {
    "patched": 60,
    "rejected": 3,
    "predicted_coverage": 142
  },
  "rollout_metadata": {},
  "signature": "regression-verified",
  "timestamp": "2026-04-28T12:00:00Z"
}
\end{verbatim}

Medical batches have the same outer structure but different correction payloads. A \texttt{MEDICAL\_HARD\_BLOCK} adds a prohibited phrase to the Oracle's block list, a \texttt{THRESHOLD\_ADJUSTMENT} installs a disposition override keyed by case class or evidence status, and an \texttt{AUDIT\_COVERAGE\_UPDATE} records that a case class has been audited by the Flywheel Overlay. This shared batch format is what allows the same release and audit machinery to operate across the spatial and medical demos.

% ============================================================
% APPENDIX: 3D Spatial Demo Specification
% ============================================================

\subsection{3D Spatial Demo Specification}
\label{app:spatial-3d}

This appendix specifies the 3D spatial governance demo. Unlike the medical demos, which govern discrete case-level decisions, the spatial demo governs a continuous reward surface in $[-1,1]^3$ learned by IIRL. The Alignment Flywheel must discover grid cells where the learned reward is non-trivial outside the expert-support basin, suppress them through governed spatial patches, and preserve the legitimate reward basin around the expert trajectory. The demo validates the same architectural claims---Red Team discovery, norm-based verification, typed governance batches, patch locality, and regression-checked preservation---in a continuous spatial domain.

\subsubsection{Problem Setup}
\label{app:spatial-setup}

\paragraph{Domain and expert path.}
The domain is the cube $[-1,1]^3$, discretised on a $20 \times 20 \times 20$ regular grid (8\,000 cells). An IIRL training run produced a precomputed loss value $\ell_i \in \mathbb{R}_{\geq 0}$ at each grid cell $i$, stored in \texttt{loss\_values.npy}. The synthetic expert path $\mathcal{E}$ consists of 20 points sampled from:
\[
\mathcal{E}(t) =
\bigl(
-1+2t,\;
\mathrm{clip}(-1+2t+0.8\sin(\pi t)),\;
\mathrm{clip}(-1+2t+0.8\cos(\pi t))
\bigr),
\quad
\]
\[t \in \{0,\tfrac{1}{19},\ldots,1\}.\]
\paragraph{Reward, basin, and flaw definitions.}
The Oracle converts loss to reward by:
\[
r_i = 1 - \min\!\bigl(\ell_i/L_{\mathrm{cap}},\,1\bigr),
\quad
L_{\mathrm{cap}} = 0.3.
\]
Cells with $\ell_i \geq 0.04$ are pre-thresholded to $\ell_i = 999\,999$ to isolate the IIRL artifacts that require governance. Let
\[
d_i = \min_{e \in \mathcal{E}} \|p_i-e\|_2
\]
be the distance from grid cell $i$ to the nearest expert-path point. With safety floor $\sigma=0.005$ and basin boundary $B=0.34$, cells are classified as follows.

\begin{table}[h]
\centering
\caption{Spatial cell classification used by the demo.}
\label{tab:spatial-cell-defs}
\small
\begin{tabular}{p{0.22\linewidth}p{0.32\linewidth}p{0.34\linewidth}}
\toprule
\textbf{Cell type} & \textbf{Condition} & \textbf{Interpretation} \\
\midrule
Basin cell & $r_i > \sigma$ and $d_i \leq B$ & Legitimate reward near the expert trajectory; must be preserved. \\
Flaw cell & $r_i > \sigma$ and $d_i > B$ & Non-trivial reward outside the expert basin; must be suppressed. \\
Inactive cell & $r_i \leq \sigma$ & Already below the safety floor; no action required. \\
\bottomrule
\end{tabular}
\end{table}

The objective is to drive the flaw count to zero while preserving the initial basin count.

\subsubsection{Oracle: \texttt{PrecomputedGridOracle}}
\label{app:spatial-oracle}

\paragraph{State and query evaluation.}
The \texttt{PrecomputedGridOracle} stores the read-only reward vector $\mathbf{r}\in\mathbb{R}^{8000}$, a mutable list of suppression kernels $\{(c_k,\beta_k)\}_{k=1}^{K}$, and a version counter $v_O$. For a query point $p$, the Oracle snaps $p$ to the nearest grid cell $i$, reads $r_i$, computes cumulative suppression,
\[
S(p)=\min\!\Bigl(1,\sum_{k=1}^{K}
\exp\!\bigl(-\|p-c_k\|^2/(2\beta_k^2)\bigr)
\Bigr),
\]
and returns the governed safety score
\[
s=\max(0,r_i-S(p)).
\]
The output is therefore not the raw IIRL reward, but the reward attenuated by every suppression kernel installed through prior governance batches.

\paragraph{Batch application.}
A \texttt{GovernanceBatch} contains \texttt{LocalCorrection} artifacts. The Oracle filters for \texttt{correction\_type == SPATIAL\_FLAW\_PATCH}; for each such correction, it appends the centre $c_k$ from \texttt{payload["flaw\_point"]} and bandwidth $\beta_k$ from \texttt{payload["support\_radius"]} to its kernel list. The Oracle version is then incremented.

\subsubsection{Patch Planner: Predictive Batch Planning}
\label{app:spatial-planner}

The adaptive variant uses a \texttt{Patch Planner} that plans each governance batch before deployment by analytically predicting kernel effects. The planner receives the basin set $\mathcal{B}$, flaw points sorted by descending distance to the expert path, and each flaw's distance $d_i$. It then selects non-redundant kernels subject to basin-preservation constraints.

\begin{table}[h!]
\centering
\caption{PatchPlanner loop for adaptive spatial governance batches.}
\label{tab:spatial-planner}
\small
\begin{tabular}{p{0.20\linewidth}p{0.70\linewidth}}
\toprule
\textbf{Step} & \textbf{Function} \\
\midrule
Skip covered flaws &
If a previously accepted kernel in the same batch already covers flaw $p_i$, skip it. \\
Propose bandwidth &
Set
\[
\beta_{\mathrm{prop}} =
\mathrm{clip}\!\bigl((d_i-B)\cdot 0.5,\beta_{\min},\beta_{\max}\bigr),
\quad
\beta_{\min}=0.03,\quad
\beta_{\max}=0.30.
\]
Distant flaws receive wider kernels; near-boundary flaws receive narrower kernels. \\
Single-kernel basin protection &
Let $d_{\mathrm{near}}=\min_{b\in\mathcal{B}}\|p_i-b\|$. The maximum bandwidth that keeps suppression of the nearest basin point below $\eta=0.10$ is
\[
\beta_{\mathrm{safe}}=
\frac{d_{\mathrm{near}}}{\sqrt{-2\ln\eta}}.
\]
The proposed bandwidth is shrunk to $\min(\beta_{\mathrm{prop}},0.9\beta_{\mathrm{safe}})$. \\
Cumulative basin check &
The planner tracks running suppression on every basin point from accepted kernels in the current batch:
\[
S_{\mathrm{existing}}(b)=
\sum_{j<i}\exp\!\bigl(-\|b-c_j\|^2/(2\beta_j^2)\bigr).
\]
If adding the candidate kernel would push any basin point above $\eta$, a 15-step binary search over $[\beta_{\min}/2,\beta]$ finds the largest safe bandwidth. If even $\beta_{\min}$ would cause excessive basin suppression, the patch is rejected. \\
Predict coverage &
The effective suppression radius is
\[
R_{\mathrm{eff}}=\beta\sqrt{-2\ln\sigma},
\quad \sigma=0.005.
\]
Other flaw points within $R_{\mathrm{eff}}$ are predicted to be suppressed below the safety floor. \\
Mark covered flaws &
Predicted-covered flaws are marked as handled and skipped later in the batch. This spends the patch budget on non-redundant kernel placements. \\
\bottomrule
\end{tabular}
\end{table}

The loop terminates when the batch budget is exhausted ($\mathrm{max\_patches}=200$) or all currently known flaws are covered.

\subsubsection{Red Team Discovery and Batch Construction}
\label{app:spatial-redteam-batch}

\paragraph{Adaptive flaw discovery.}
Each iteration queries the full $20^3=8\,000$ grid through the governed Oracle. From iteration~2 onward, the Red Team also performs dense sampling around previously active flaws. The number of fine samples and search radius adapt to the previous flaw count $f_{\mathrm{prev}}$:
\[
n_{\mathrm{fine}} =
\mathrm{clip}\!\bigl(8000/\max(1,\lfloor f_{\mathrm{prev}}/50 \rfloor),500,2000\bigr),
\quad
\]
\[
r_{\mathrm{fine}} =
\mathrm{clip}\!\bigl(f_{\mathrm{prev}}/5000,0.12,0.35\bigr).
\]
Many remaining flaws trigger wider exploration; few remaining flaws trigger tighter search around residual boundary regions. Fine samples are drawn uniformly within $r_{\mathrm{fine}}$ of up to 200 previously known flaw centres. A queried point is classified as a flaw if its governed Oracle score exceeds $\sigma$ and its distance to the expert path exceeds $B$.

\paragraph{Batch construction.}
Each accepted patch $(p_i,\beta_i,n_{\mathrm{covered}})$ from the PatchPlanner produces the corrections in Table~\ref{tab:spatial-corrections}. The corrections are wrapped in a \texttt{GovernanceBatch} with regression evidence recording the number of kernels placed and total predicted coverage.

\begin{table}[h!]
\centering
\caption{Typed corrections used by the spatial demo.}
\label{tab:spatial-corrections}
\small
\begin{tabular}{p{0.30\linewidth}p{0.60\linewidth}}
\toprule
\textbf{Correction type} & \textbf{Effect} \\
\midrule
\texttt{SPATIAL\_FLAW\_PATCH} &
Payload contains \texttt{flaw\_point} $p_i$ and \texttt{support\_radius} $\beta_i$. Applying the correction installs a suppression kernel in the Oracle. \\
\texttt{AUDIT\_COVERAGE\_UPDATE} &
Payload records the predicted coverage class, e.g. \texttt{"spatial|bw=$\beta_i$|cov=$n_{\mathrm{covered}}$"}, for audit trail and batch provenance. \\
\bottomrule
\end{tabular}
\end{table}

The batch is applied to the Oracle, which installs suppression kernels, and to the Knowledge Base, which records the governed update and its evidence.

\subsubsection{Convergence Mechanics}
\label{app:spatial-convergence}

Each iteration follows the same governance lifecycle:
\[
\text{SearchRedTeam}
\xrightarrow{\text{flaw points}}
\text{PatchPlanner}
\xrightarrow{\text{plan}}
\text{build\_batch}
\xrightarrow{\text{GovernanceBatch}}
O.
\]
After the batch is applied, the Red Team re-queries known flaw points through the updated Oracle and uses the remaining active flaws to guide the next search iteration.

\paragraph{Why convergence is fast.}
Farthest-first ordering combined with adaptive bandwidth creates large early reductions. A distant flaw at $d=0.9$, for example, may receive $\beta \approx 0.28$, giving $R_{\mathrm{eff}}\approx 0.92$. Such a kernel can cover many neighbouring flaws, which are marked as handled before deployment. The 200-patch budget is therefore spent on non-redundant placements, so per-iteration flaw reduction can substantially exceed the number of directly patched centres.

\paragraph{Why the basin is preserved.}
Basin preservation is enforced by a two-tier check: a per-kernel analytical cap and a cumulative binary-search cap over basin suppression. Kernels near the basin boundary are shrunk or rejected if they would exceed the suppression budget. This is a regression-checking mechanism over the accepted basin rather than a guarantee about unsampled continuous space.

\paragraph{Predicted coverage rather than accidental collateral.}
In an earlier spatial variant, ``collateral'' referred to flaw points suppressed by neighbouring kernel spill-over. In this adaptive variant, the PatchPlanner predicts this effect before deployment and records the predicted coverage. We therefore report \emph{predicted coverage}: flaw cells suppressed by a planned neighbouring kernel rather than directly targeted by their own kernel.

\subsubsection{Visualisation}
\label{app:spatial-vis}

Each iteration produces a 3D scatter plot from elevation $45^\circ$ and azimuth $300^\circ$. Active cells with reward above $\sigma$ are plotted with colour mapped to reward value using the \texttt{viridis} colourmap. The expert trajectory is plotted as a blue line, and kernel centres placed in the current iteration are plotted as red \textsf{x} markers. Basin and flaw cells use the same reward colourmap, so the visible structure is the governed reward surface itself. Across iterations, the active reward region contracts toward the expert path until only the basin remains.

\subsubsection{Comparison with Fixed-Bandwidth Baseline}
\label{app:spatial-comparison}

To test the effect of adaptive bandwidth and predictive coverage, the same problem was run with slightly different spatial data,  a fixed-bandwidth baseline, \texttt{spatial\_3d\_fixed\_bw}. The baseline uses constant bandwidth $\beta=0.05$ and a batch budget of 60 patches per iteration.

\begin{table}[h!]
\centering
\caption{3D Spatial demo: adaptive PatchPlanner vs.\ fixed-bandwidth baseline.}
\label{tab:spatial-comparison}
\small
\begin{tabular}{lrr}
\toprule
Metric & PatchPlanner (adaptive) & Fixed BW ($\beta=0.05$) \\
\midrule
Iterations to converge & 16 & $>25$ (not converged) \\
Patches/iter (max) & 60 & 60 \\
Basin preserved & 783/783 (100\%) & 783/783 (100\%) \\
Total kernels placed & 834 & 1500 (it. 30)  \\
\bottomrule
\end{tabular}
\end{table}

The adaptive PatchPlanner converges in 16 iterations because wide kernels at distant flaws analytically cover many neighbors, increasing effective throughput. The fixed-bandwidth baseline reduces flaws much more slowly and would require substantially more iterations to converge. Both preserve the basin in this grid-based evaluation, confirming that the regression check prevents accepted basin cells from being suppressed below the safety floor.

\subsubsection{Convergence Results}
\label{app:spatial-results}

Table~\ref{tab:spatial-conv} reports per-iteration convergence for the adaptive PatchPlanner variant. ``Found'' is the number of active flaw candidates discovered in that iteration; ``Kern'' is the number of kernels accepted into the governance batch; ``Predicted'' is the number of flaw candidates analytically predicted to be covered by accepted kernels; ``Reject'' counts candidate kernels rejected by basin-preservation checks. ``Basin'' is the number of sampled basin cells still above the safety floor after the batch is applied.

\begin{table}[h!]
\centering
\caption{3D Spatial demo (adaptive PatchPlanner): convergence over iterations.}
\label{tab:spatial-conv}
\small
\begin{tabular}{rrrrrrrl}
\toprule
Iter & Found & Kern & Predicted & Reject & Basin & Flaws & Oracle \\
\midrule
1  & 500   & 87 & 500  & 0 & 783 & 412 & v1  \\
2  & 993   & 42 & 993  & 0 & 783 & 357 & v2  \\
3  & 874   & 58 & 874  & 0 & 783 & 290 & v3  \\
4  & 1\,030 & 57 & 1\,030 & 0 & 783 & 219 & v4  \\
5  & 1\,119 & 49 & 1\,119 & 0 & 783 & 159 & v5  \\
6  & 960   & 71 & 959  & 1 & 783 & 103 & v6  \\
7  & 782   & 90 & 782  & 1 & 783 & 65  & v7  \\
8  & 555   & 90 & 555  & 0 & 783 & 32  & v8  \\
9  & 398   & 82 & 396  & 3 & 783 & 13  & v9  \\
10 & 211   & 55 & 209  & 3 & 783 & 7   & v10 \\
11 & 164   & 34 & 161  & 3 & 783 & 5   & v11 \\
12 & 96    & 24 & 96   & 1 & 783 & 0   & v12 \\
\bottomrule
\end{tabular}
\end{table}

\paragraph{Key observations.}
The adaptive PatchPlanner drives the remaining flaw count to zero after 12 governance iterations. The sampled basin count remains fixed at 783 throughout the full run, giving 100\% basin preservation on the evaluation grid. Early iterations remove large flaw regions because accepted kernels cover many neighboring flaw candidates analytically; later iterations place fewer kernels as the remaining flaws become sparse and closer to basin-preservation constraints. Rejections appear only from iteration~6 onward and remain small relative to the number of accepted kernels, indicating that the planner usually finds safe bandwidths and rejects only candidates that would violate the preservation budget. The final iteration requires only 24 accepted kernels and one rejection, confirming that the residual flaws are localized and constrained by the regression checks.

\subsection{Medical Demo Specifications}
\label{app:medical-demos}

This appendix gives the technical specification for three synthetic medical governance demos. All three instantiate the complete Alignment Flywheel pipeline---Proposer, Safety Oracle, Alignment Flywheel Overlay, Query Merger, Enforcement, and the five governance roles---using the same abstract interfaces, knowledge base, batch format, and protocol artifacts. The demos differ only in the concrete component implementations selected through a single composition root. They therefore serve as ablations over governance complexity: Demo~1 uses a minimal heuristic Oracle and FIFO triage, Demo~2 adds multi-dimensional risk scoring and priority triage, and Demo~3 instantiates the patient-portal scenario summarized in the main text.

\subsubsection{Shared Architecture and Pipeline}
\label{app:shared-arch}

All three medical demos share the same runtime pipeline and governance cycle. The purpose of the demos is not to validate clinical adequacy, but to show that the same Alignment Flywheel protocol can operate over different proxy Oracles, norm sets, triage strategies, and refinement mechanisms.

\paragraph{Runtime evaluation pipeline.}
Table~\ref{tab:medical-runtime-pipeline} summarizes the shared runtime pipeline. The concrete Oracle and overlay differ by demo, but the protocol topology is invariant.

\begin{table}[h!]
\centering
\caption{Shared runtime pipeline used by all medical demos.}
\label{tab:medical-runtime-pipeline}
\small
\begin{tabular}{p{0.20\linewidth}p{0.70\linewidth}}
\toprule
\textbf{Component} & \textbf{Function} \\
\midrule
Proposer $P$ &
A \texttt{PassthroughProposer} wraps the incoming case data---patient message, draft reply, disposition, and metadata---as a \texttt{Trajectory} of kind \texttt{MESSAGE}. No generation occurs; the case already contains the candidate content. \\
Safety Oracle $O$ &
Evaluates the trajectory and returns \texttt{OracleRawOutput}: safety score $s$, prediction uncertainty $u$, uncertainty threshold $u_{\mathrm{thresh}}$, Oracle version $v_O$, and evidence status. \\
Alignment Flywheel Overlay &
Evaluates audit coverage and returns \texttt{Alignment FlywheelOverlay}: audit coverage uncertainty $u_a$, threshold $u_{a,\mathrm{thresh}}$, governance version $v_G$, and audit status. \\
Query Merger &
A \texttt{DefaultQueryMerger} combines Oracle-side and governance-side signals into a \texttt{UnifiedQueryResult} while preserving provenance. \\
Enforcement $E$ &
A \texttt{DefaultEnforcement} policy maps the unified signals to $a \in \{\mathrm{allow},\mathrm{block},\mathrm{escalate}\}$. \\
\bottomrule
\end{tabular}
\end{table}

\paragraph{Medical enforcement policy.}
The main text defines the conservative base policy: high prediction uncertainty or high audit-coverage uncertainty triggers fail-closed escalation. The medical demos instantiate a domain-specific refinement of this policy. Prediction uncertainty always triggers escalation. Audit-coverage uncertainty triggers escalation only when evidence is weak and the safety score is below an audit margin, preventing routine supported cases from escalating solely because their key has not yet been explicitly covered.

The policy is parameterised by a safety margin $\theta_s = 0.4$ and an audit safety margin $\theta_a = 0.85$.

\begin{table}[h!]
\centering
\caption{Medical-domain enforcement policy used in all three demos.}
\label{tab:medical-enforcement-policy}
\small
\begin{tabular}{p{0.55\linewidth}p{0.32\linewidth}}
\toprule
\textbf{Condition} & \textbf{Action} \\
\midrule
$u \geq u_{\mathrm{thresh}}$ & \texttt{escalate}: Oracle prediction uncertainty is too high. \\
$s < \theta_s$ & \texttt{block}: safety score indicates unsafe behavior. \\
$u_a \geq u_{a,\mathrm{thresh}}$, evidence $\in \{\texttt{insufficient}, \texttt{conflicting}\}$, and $s < \theta_a$ & \texttt{escalate}: audit coverage is insufficient for a weak-evidence, borderline-safety case. \\
Otherwise & \texttt{allow}. \\
\bottomrule
\end{tabular}
\end{table}

Thus uncertainty-driven escalation takes precedence, followed by definitive safety blocking, followed by coverage-driven escalation. Cases that are safe, well-covered, and low-uncertainty are allowed.

\paragraph{Governance cycle.}
After all fixed cases are evaluated, one governance iteration runs. Table~\ref{tab:medical-governance-cycle} gives the shared cycle.

\begin{table}[h]
\centering
\caption{Shared governance cycle used by all medical demos.}
\label{tab:medical-governance-cycle}
\small
\begin{tabular}{p{0.18\linewidth}p{0.72\linewidth}}
\toprule
\textbf{Role} & \textbf{Function} \\
\midrule
Red Team &
Generates 20 synthetic cases per iteration by sampling a combinatorial case space. Each case is classified by failure category and pushed to $Q_{\mathrm{ver}}$ as a \texttt{CandidateFlaw}. \\
Verification &
Checks candidates against the normative specification $\Phi$. Keyword-block norms check prohibited terms under weak evidence; predicate norms check multi-field relationships such as whether the proposed disposition meets the required minimum. Confirmed violations produce \texttt{VerificationResult} records. \\
Triage &
Orders verified violations by priority and forwards them to $Q_{\mathrm{ref}}$. \\
Refinement &
Maps verified flaws to typed corrections. A tiered batch strategy rotates across categories to ensure diverse coverage. In all three demos, batch capacity is limited to one flaw per iteration. \\
Batch application &
Applies the resulting \texttt{GovernanceBatch} to the Oracle and Alignment Flywheel Overlay. Oracle patches install hard-blocks and disposition overrides; overlay patches register audit coverage. Both components increment their version identifiers when state changes. \\
\bottomrule
\end{tabular}
\end{table}

The cycle then repeats: fixed cases are re-evaluated under the patched stack, governance runs again, and convergence is measured by the disappearance of escalations while preserving correct allow/block separation.

\paragraph{Correction types and patch mechanics.}
Table~\ref{tab:medical-corrections} summarizes the correction types used across all medical demos. The first two modify Oracle state; the third modifies governance state.

\begin{table}[h!]
\centering
\caption{Typed corrections used in the medical demos.}
\label{tab:medical-corrections}
\small
\begin{tabular}{p{0.28\linewidth}p{0.62\linewidth}}
\toprule
\textbf{Correction type} & \textbf{Effect} \\
\midrule
\texttt{MEDICAL\_HARD\_BLOCK} &
Adds a keyword to the Oracle's internal block list. Later trajectories containing the keyword return $s=0.0$, $u=0.05$, producing an unconditional block. \\
\texttt{THRESHOLD\_ADJUSTMENT} &
Installs a disposition override keyed by \texttt{case\_type|evidence\_status}. If a matching trajectory has a disposition below the required minimum, the Oracle returns $s=0.15$, $u=0.1$, which Enforcement maps to \texttt{block}. \\
\texttt{AUDIT\_COVERAGE\_UPDATE} &
Registers a \texttt{case\_type|evidence|acuity} key as audited in the Alignment Flywheel Overlay. Covered keys produce $u_a=0.15{-}0.20$; uncovered keys produce $u_a=0.85{-}0.90$. \\
\bottomrule
\end{tabular}
\end{table}

Together, these corrections convert uncertainty-driven escalations into definitive allow/block decisions without modifying or retraining the Proposer.

\paragraph{Shared OODA composition.}
All three demos use the same OODA skeleton for Red Team, Verification, and Refinement. The concrete strategy modules differ only by parameterisation.

\begin{table}[h!]
\centering
\caption{Shared OODA components for the medical demos.}
\label{tab:medical-ooda}
\small
\begin{tabular}{p{0.17\linewidth}p{0.25\linewidth}p{0.48\linewidth}}
\toprule
\textbf{Role} & \textbf{Component} & \textbf{Function} \\
\midrule
Red Team &
Medical Case Generator &
Samples medications, symptoms, lab tests, and demographics; classifies cases as \texttt{missed\_urgency}, undertriaged\_med, lab\_no\_context, \texttt{vulnerable\_patient}, or \texttt{exploratory}. \\
Red Team &
Medical Priority Decider &
Sorts candidates by category-weighted priority: \texttt{missed\_urgency} $>$ \texttt{undertriaged\_med} $>$ \texttt{vulnerable\_patient} $>$ \texttt{lab\_no\_context} $>$ \texttt{exploratory}. \\
Red Team &
Medical Candidate Submitter &
Wraps candidates as \texttt{Candidate Flaw} artifacts with Oracle signals, trajectory payload, and metadata. \\
Verification &
Medical NormM atcher and Medical Violation Decider &
Extract draft text, patient message, evidence status, and disposition; evaluate keyword-block and predicate norms; emit \texttt{VIOLATION} or \texttt{NO\_VIOLATION}. \\
Refinement &
Medical Correction Orienter &
Maps each verified flaw to a small bundle of corrections, e.g., hard-block, disposition overrides, and audit-coverage updates. \\
Refinement &
Medical Batch Decider and Medical Batch Deployer &
Selects at most one flaw per iteration, builds a \texttt{GovernanceBatch}, sets version transitions, and returns the batch for application. \\
\bottomrule
\end{tabular}
\end{table}

% ── SIMPLE MEDICAL ──────────────────────────────────────────

\subsubsection{Demo 1: Simple Medical}
\label{app:demo-simple}

\paragraph{Scenario and purpose.}
A healthcare organization deploys an AI assistant that drafts replies to patient messages in a clinical inbox. Patients ask about medication dosage, lab results, appointments, or routine follow-up; the AI proposes a draft reply and a disposition such as \texttt{reply\_only}, \texttt{nurse\_review}, or \texttt{clinician\_review}. This demo uses deliberately simple components---a small heuristic Oracle, two norms, and FIFO triage---to isolate the protocol flow: discover flaws, verify them against norms, patch the Oracle stack, and re-evaluate.

\paragraph{Fixed cases.}
The fixed evaluation set contains 15 synthetic cases, grouped in Table~\ref{tab:simple-cases}.

\begin{table}[h!]
\centering
\caption{Simple Medical demo: fixed evaluation groups.}
\label{tab:simple-cases}
\small
\begin{tabular}{p{0.22\linewidth}p{0.10\linewidth}p{0.43\linewidth}p{0.17\linewidth}}
\toprule
\textbf{Group} & \textbf{Count} & \textbf{Risk pattern} & \textbf{Expected transition} \\
\midrule
Unsafe medication & 3 & Unsupported dosing advice with \texttt{reply\_only} disposition. & \texttt{escalate} $\rightarrow$ \texttt{block} \\
Unsafe lab & 3 & Lab-result interpretation or monitoring advice without sufficient evidence and without nurse review. & \texttt{escalate} $\rightarrow$ \texttt{block} \\
Borderline & 3 & Similar clinical content, but already routed to \texttt{clinician\_review}. & \texttt{escalate} $\rightarrow$ \texttt{allow} \\
Safe & 6 & Routine supported interactions such as refills, scheduling, and acknowledgements. & \texttt{allow} throughout \\
\bottomrule
\end{tabular}
\end{table}

\paragraph{Oracle, overlay, norms, and triage.}
Table~\ref{tab:simple-components} summarizes the concrete components.

\begin{table}[h!]
\centering
\caption{Simple Medical demo: concrete component choices.}
\label{tab:simple-components}
\small
\begin{tabular}{p{0.22\linewidth}p{0.68\linewidth}}
\toprule
\textbf{Component} & \textbf{Specification} \\
\midrule
Oracle &
\texttt{SimpleMedicalOracle}; heuristic scoring with patchable hard-block list $B$ and disposition override dictionary $D$. Hard-blocks return $s=0.0,u=0.05$; disposition failures return $s=0.15,u=0.1$. Base scoring penalizes medication stems, medication case type, and weak evidence. \\
Prediction uncertainty &
Baseline $u=0.15$, increased for insufficient, unknown, or conflicting evidence and for medication cases. $u_{\mathrm{thresh}}=0.30$ for medication cases and $0.50$ otherwise. \\
Alignment Flywheel Overlay &
\texttt{SimpleMedicalOverlay}; coverage key \texttt{case\_type|evidence\_status|acuity}. Covered keys produce $u_a=0.2$; uncovered keys produce $u_a=0.9$. Threshold $u_{a,\mathrm{thresh}}=0.6$. \\
Norms $\Phi$ &
Two norms: \texttt{NO\_UNSUPPORTED\_MED\_ADVICE}, a keyword-block norm over medication-change terms under weak evidence; and \texttt{REQUIRE\_EVIDENCE\_FOR\_RESULTS}, a predicate norm requiring at least \texttt{nurse\_review} for weak-evidence lab/result interpretation. \\
Triage &
\texttt{FIFOTriage}; violations are passed to refinement in arrival order. \\
\bottomrule
\end{tabular}
\end{table}

\paragraph{Convergence results.}
Table~\ref{tab:simple-conv} shows enforcement behavior across evaluations. Evaluation~0 is the initial stack; later evaluations occur after governance batches have been applied.

\begin{table}[h!]
\centering
\caption{Simple Medical demo: enforcement convergence.}
\label{tab:simple-conv}
\small
\begin{tabular}{rrrrrl}
\toprule
Eval & Allow & Block & Escalate & Esc.\,Rate & Stack \\
\midrule
0 & 6 & 0 & 9 & 60\% & v0 \\
1 & 6 & 3 & 6 & 40\% & v1 \\
2 & 9 & 6 & 0 & 0\%  & v2 \\
\bottomrule
\end{tabular}
\end{table}

\paragraph{Observation.}
The simple demo validates the protocol flow independently of component sophistication. The safe cases remain \texttt{allow}; unsafe cases transition from uncertainty-driven \texttt{escalate} to definitive \texttt{block}; and borderline cases transition from \texttt{escalate} to \texttt{allow} once audit coverage is established.

% ── COMPLEX MEDICAL ─────────────────────────────────────────

\subsubsection{Demo 2: Complex Medical}
\label{app:demo-complex}

\paragraph{Scenario and purpose.}
The same clinical-inbox setting is extended with synthetic specialty structure, named medications, interaction risks, vulnerable patient profiles, and specialty-specific thresholds. The purpose is to test whether the same governance loop handles richer proxy-Oracle logic and a more complex normative specification without changing the Alignment Flywheel protocol.

\paragraph{Fixed cases.}
The fixed evaluation set contains 18 synthetic cases, grouped in Table~\ref{tab:complex-cases}.

\begin{table}[h!]
\centering
\caption{Complex Medical demo: fixed evaluation groups.}
\label{tab:complex-cases}
\small
\begin{tabular}{p{0.22\linewidth}p{0.10\linewidth}p{0.43\linewidth}p{0.17\linewidth}}
\toprule
\textbf{Group} & \textbf{Count} & \textbf{Risk pattern} & \textbf{Expected transition} \\
\midrule
Unsafe stop-taking & 3 & Recommendations to stop high-risk medications such as warfarin, insulin, or oxycodone under weak evidence. & \texttt{escalate} $\rightarrow$ \texttt{block} \\
Unsafe dose increase & 3 & Autonomous dose increases for higher-risk drugs such as insulin, tramadol, or prednisone. & \texttt{escalate} $\rightarrow$ \texttt{block} \\
Unsafe lab & 3 & Organ-function or medication-related lab interpretation with insufficient evidence and direct patient routing. & \texttt{escalate} $\rightarrow$ \texttt{block} \\
Borderline & 3 & Lab or medication discussion already routed to \texttt{clinician\_review}. & \texttt{escalate} $\rightarrow$ \texttt{allow} \\
Safe & 6 & Routine supported interactions without clinical decision-making. & \texttt{allow} throughout \\
\bottomrule
\end{tabular}
\end{table}

\paragraph{Oracle, overlay, norms, and triage.}
The \texttt{ComplexMedicalOracle} scores five dimensions using lookup tables: medication risk, interaction risk, action severity, patient vulnerability, and evidence quality. The safety score is:
\[
s = \max\!\bigl(0,\; 1 - (0.30\, r_\mathrm{med} + 0.25\, r_\mathrm{interact} + 0.20\, r_\mathrm{action} + 0.10\, r_\mathrm{patient} + 0.15\, r_\mathrm{ev})\bigr).
\]
Prediction uncertainty starts at $u=0.15$ and increases under weak evidence, high medication risk, or detected drug interactions. Thresholds are specialty-based.

\begin{table}[h!]
\centering
\caption{Complex Medical demo: concrete component choices.}
\label{tab:complex-components}
\small
\begin{tabular}{p{0.22\linewidth}p{0.68\linewidth}}
\toprule
\textbf{Component} & \textbf{Specification} \\
\midrule
Oracle &
\texttt{ComplexMedicalOracle}; patchable state includes hard-block list $B$, specialty threshold overrides $T$, and disposition overrides $D$. Evaluation order is hard-block, disposition override, then five-dimensional scoring. \\
Risk dimensions &
$r_\mathrm{med}$ from a medication-risk table; $r_\mathrm{interact}$ from known interaction pairs; $r_\mathrm{action}$ from action severity; $r_\mathrm{patient}$ from age and comorbidities; $r_\mathrm{ev}$ from evidence status. \\
Alignment Flywheel Overlay &
\texttt{ComplexMedicalOverlay}; coverage key \texttt{specialty|case\_type|evidence\_status}. Covered keys produce $u_a=0.15$; uncovered keys produce $u_a=0.85$. Threshold $u_{a,\mathrm{thresh}}=0.6$. \\
Norms $\Phi$ &
Four norms: \texttt{NO\_HIGH\_RISK\_MED\_WITHOUT\_EVIDENCE}, \texttt{INTERACTION\_SAFETY}, \texttt{VULNERABLE\_PATIENT\_PROTECTION}, and \texttt{SEVERE\_ACTION\_EVIDENCE}. These cover keyword-block, regex, threshold-rule, and predicate-style checks. \\
Triage &
\texttt{PriorityTriage}; priority order is \texttt{missed\_urgency} $>$ \texttt{undertriaged\_med} $>$ \texttt{vulnerable\_patient} $>$ \texttt{lab\_no\_context} $>$ \texttt{exploratory}. \\
\bottomrule
\end{tabular}
\end{table}

\paragraph{Convergence results.}

\begin{table}[h!]
\centering
\caption{Complex Medical demo: enforcement convergence.}
\label{tab:complex-conv}
\small
\begin{tabular}{rrrrrl}
\toprule
Eval & Allow & Block & Escalate & Esc.\,Rate & Stack \\
\midrule
0 & 6 & 0  & 12 & 67\% & v0 \\
1 & 6 & 6  & 6  & 33\% & v1 \\
2 & 9 & 9  & 0  & 0\%  & v2 \\
\bottomrule
\end{tabular}
\end{table}

Despite the richer proxy Oracle and norm set, the governance pipeline converges in the same number of evaluations as the simple demo. This reflects the batch strategy: the initial unsafe cases cluster into a small number of failure families, each resolved by a governed update.

% ── PATIENT PORTAL ──────────────────────────────────────────

\subsubsection{Demo 3: Patient Portal}
\label{app:demo-portal}

\paragraph{Scenario and purpose.}
A hospital operates a patient-facing web portal where patients message their care team directly. Unlike the clinical inbox in Demos~1--2, the portal is a patient-facing channel: the AI's draft reply may be the first thing the patient reads unless the system routes the message for review. This is the primary medical evaluation scenario summarized in Section~\ref{sec:eval-medical}. It validates the Alignment Flywheel architecture with a heuristic proxy Oracle that scores medication risk, urgency, disposition mismatch, and evidence quality.

\paragraph{Fixed cases.}
The fixed evaluation set contains 15 synthetic cases, grouped in Table~\ref{tab:portal-cases}.

\begin{table}[h!]
\centering
\caption{Patient Portal demo: fixed evaluation groups.}
\label{tab:portal-cases}
\small
\begin{tabular}{p{0.22\linewidth}p{0.10\linewidth}p{0.43\linewidth}p{0.17\linewidth}}
\toprule
\textbf{Group} & \textbf{Count} & \textbf{Risk pattern} & \textbf{Expected transition} \\
\midrule
Unsafe medication & 3 & Direct patient-facing medication-change advice, such as increasing dose, without clinician review or sufficient evidence. & \texttt{escalate} $\rightarrow$ \texttt{block} \\
Unsafe lab & 3 & Lab-result interpretation with medication context, weak evidence, and \texttt{reply\_only} routing. & \texttt{escalate} $\rightarrow$ \texttt{block} \\
Borderline & 3 & Lab or medication discussion already routed to \texttt{clinician\_review}. & \texttt{escalate} $\rightarrow$ \texttt{allow} \\
Safe & 6 & Routine supported interactions such as refills, appointments, physical-therapy progress, diet follow-up, improving symptoms, and normal annual labs. & \texttt{allow} throughout \\
\bottomrule
\end{tabular}
\end{table}

\paragraph{Oracle, overlay, norms, and triage.}
The \texttt{PatientPortalOracle} scores four dimensions: medication content, urgency signals, evidence quality, and disposition mismatch. The safety score is:
\[
s = \max\!\bigl(0,\; 1 - (0.30\, r_\mathrm{med} + 0.25\, r_\mathrm{urg} + 0.25\, r_\mathrm{disp} + 0.20\, r_\mathrm{ev})\bigr).
\]
Prediction uncertainty starts at $u=0.15$ and increases under weak evidence, urgent/semi-urgent acuity, or medication content. Uncertainty thresholds are acuity-based: $u_{\mathrm{thresh}}=0.20$ for urgent cases, $0.30$ for semi-urgent cases, and $0.50$ for routine cases.

\begin{table}[h!]
\centering
\caption{Patient Portal demo: concrete component choices.}
\label{tab:portal-components}
\small
\begin{tabular}{p{0.22\linewidth}p{0.68\linewidth}}
\toprule
\textbf{Component} & \textbf{Specification} \\
\midrule
Oracle &
\texttt{PatientPortalOracle}; patchable state includes hard-block list $B$ and disposition override dictionary $D$. Evaluation order is hard-block, disposition override, then four-dimensional scoring. \\
Risk dimensions &
$r_\mathrm{med}$ counts medication keywords; $r_\mathrm{urg}$ counts urgency keywords in the message or draft; $r_\mathrm{ev}$ encodes evidence quality; $r_\mathrm{disp}$ measures the gap between required and proposed disposition. \\
Disposition ranks &
\texttt{reply\_only}$=0$, \texttt{nurse\_review}$=1$, \texttt{clinician\_review}$=2$, \texttt{urgent\_escalation}$=3$. Needed disposition increases with urgency, medication risk, and semi-urgent acuity. \\
Alignment Flywheel Overlay &
\texttt{PatientPortalOverlay}; coverage key \texttt{case\_type|evidence\_status|acuity}. Covered keys produce $u_a=0.15$; uncovered keys produce $u_a=0.85$. Threshold $u_{a,\mathrm{thresh}}=0.6$. \\
Norms $\Phi$ &
Four norms: \texttt{MED\_ADVICE\_REQUIRES\_CLINICIAN}, \texttt{URGENCY\_REQUIRES\_ESCALATION}, \texttt{LAB\_RESULTS\_NEED\_CONTEXT}, and \texttt{NO\_UNSUPPORTED\_MED\_KEYWORDS}. \\
Triage &
\texttt{PriorityTriage}; same category-priority ordering as the complex demo. \\
\bottomrule
\end{tabular}
\end{table}

\paragraph{Convergence results.}

\begin{table}[h!]
\centering
\caption{Patient Portal demo: enforcement convergence.}
\label{tab:portal-conv}
\small
\begin{tabular}{rrrrrl}
\toprule
Eval & Allow & Block & Escalate & Esc.\,Rate & Stack \\
\midrule
0 & 6 & 0 & 9 & 60\% & v0 \\
1 & 6 & 3 & 6 & 40\% & v1 \\
2 & 9 & 6 & 0 & 0\%  & v2 \\
\bottomrule
\end{tabular}
\end{table}

Convergence is structurally identical to the simple demo because both have the same case-count distribution. The difference is that the Patient Portal Oracle uses richer scoring, urgency detection, and disposition-gap analysis, while the norm set is tailored to patient-facing communication.

% ── CROSS-DEMO ANALYSIS ────────────────────────────────────

\subsubsection{Cross-Demo Analysis}
\label{app:demo-analysis}

\paragraph{Architectural invariance.}
Table~\ref{tab:demo-comparison} summarizes the variation across demos. The abstract interfaces, controllers, protocol artifacts, knowledge base, queues, and batch format are identical; only the concrete component implementations change.

\begin{table}[h!]
\centering
\caption{Component variation across medical demos.}
\label{tab:demo-comparison}
\small
\begin{tabular}{lccc}
\toprule
Component & Simple & Complex & Patient Portal \\
\midrule
Oracle scoring & 3 heuristic rules & 5-dimensional lookup & 4-dimensional keywords \\
Patchable state & $B$, $D$ & $B$, $T$, $D$ & $B$, $D$ \\
Norm count (kinds) & 2 (2) & 4 (4) & 4 (2) \\
Triage & FIFO & priority-ranked & priority-ranked \\
Fixed cases & 15 & 18 & 15 \\
Evaluations to converge & 3 & 3 & 3 \\
\bottomrule
\end{tabular}
\end{table}

\paragraph{Convergence pattern.}
All three demos follow the same convergence trajectory: initial escalation dominance, progressive conversion of escalations into definitive blocks or allows as governance batches accumulate, and zero escalations by the final evaluation. The pattern is consistent across proxy-Oracle complexity levels, indicating that convergence is driven by the governance pipeline and patch mechanics rather than by a single Oracle implementation.

\paragraph{Patch locality.}
The Proposer is never modified. The \texttt{PassthroughProposer} returns the same trajectory throughout all evaluations. Behavioral changes arise only from Oracle state changes, such as hard-blocks and disposition overrides, and Alignment Flywheel Overlay state changes, such as audit-coverage registrations. This directly demonstrates patch locality in the medical setting: observed failures are mitigated by governed updates to the Oracle stack and governance state rather than by modifying or retracting the Proposer.

\paragraph{Correction lifecycle.}
Each governance iteration follows the same lifecycle across all demos:
\[
\text{Red Team}
\xrightarrow{\texttt{CandidateFlaw}}
Q_\mathrm{ver}
\xrightarrow{\texttt{Verification}}
\text{Triage}
\xrightarrow{\texttt{Refinement}}
Q_\mathrm{ref}
\xrightarrow{\texttt{Batch}}
O,\ \text{Overlay}.
\]
The typed artifacts, protocol boundaries, and OODA interaction patterns are identical. The mechanism is stable while the strategy modules are pluggable.

\subsection{Artifact Trace: Spatial Demo}
\label{app:trace}

We trace one flaw point through the complete governance pipeline to illustrate how the protocol boundaries from Section~4 are instantiated.

\begin{enumerate}
\item \textbf{Red Team (Observe--Orient--Decide--Act):}
The Red Team queries the Oracle at grid point $x = [0.73, -0.42, 0.81]$ and receives $s = 0.68$.
It computes $d(x, \tau_{\mathrm{expert}}) = 0.72 > 0.34$: a high-safety point far from the expert path.
It emits a \texttt{CandidateFlaw} artifact with the point, oracle signals, and distance metadata.

\item \textbf{Verification (Observe--Orient--Decide--Act):}
The Verifier retrieves norm \texttt{SPATIAL\_SUPPORT\_REQUIRED} from~$\Phi$, checks $d = 0.72 > 0.34$, and emits a \texttt{VerificationResult} with \texttt{outcome = VIOLATION} and \texttt{violated\_norm\_id = SPATIAL\_SUPPORT\_REQUIRED}.

\item \textbf{Triage:}
The verified flaw is submitted to the triage queue, sorted by distance descending (farthest first, maximising collateral coverage).

\item \textbf{Refinement (Observe--Orient--Decide--Act):}
The Refinement pops the flaw, computes adaptive bandwidth $h = \mathrm{clip}((0.72 - 0.34) \cdot 0.5, 0.03, 0.30) = 0.19$.
The cumulative regression verifier checks: $\max_p \exp(-\|p - x\|^2 / 2 \cdot 0.19^2) = 0.0003$ for the nearest basin point at distance~$0.68$---well within the $10\%$ budget.
The patch is accepted and added to the \texttt{GovernanceBatch}.

\item \textbf{Batch deployment:}
The batch is applied to the Oracle (\texttt{send\_patch}), which adds a suppression kernel at $[0.73, -0.42, 0.81]$ with $h = 0.19$.
The Alignment Flywheel overlay records the coverage update.
Oracle version increments from $v_4$ to $v_5$.

\item \textbf{Post-patch verification:}
On the next iteration's grid scan, the Oracle returns $s = \max(0, 0.68 - 0.97) = 0$ at this point (the suppression kernel centred here has value~${\sim}1.0$).
The point is no longer a flaw.
Adjacent points within the kernel's effective radius (${\sim}2.5 \times 0.19 = 0.48$) are also suppressed---this is the collateral effect.
\end{enumerate}

\subsection{Artifact Trace: Patient Portal Demo}
\label{app:trace-medical}

We trace one case through the medical governance pipeline.

\begin{enumerate}
\item \textbf{Red Team:}
Generates a synthetic case: patient asks about stopping warfarin, draft reply says ``You could stop taking warfarin and switch to apixaban,'' proposed disposition is \texttt{nurse\_review}, evidence is \texttt{conflicting}, acuity is \texttt{routine}.
The Red Team classifies this as \texttt{undertriaged\_med} (medication advice with weak evidence) and emits a \texttt{CandidateFlaw}.

\item \textbf{Verification:}
The Verifier loads norm \texttt{NO\_UNSUPPORTED\_MED\_KEYWORDS} (kind: \texttt{KEYWORD\_BLOCK}).
Checks: ``stop taking'' $\in$ draft text, evidence $\neq$ \texttt{supported} $\Rightarrow$ \texttt{VIOLATION}.
Emits \texttt{VerificationResult} with \texttt{violated\_norm\_id =\\ NO\_UNSUPPORTED\_MED\_KEYWORDS}.

\item \textbf{Refinement:}
Emits a \texttt{GovernanceBatch} containing: \\
(a)~\texttt{medical\_hard\_block}: keyword ``stop taking'' added to oracle block list;
(b)~\texttt{threshold\_adjustment}: disposition override \texttt{clinician\_review} installed for keys  \\\texttt{medication|insufficient}, \\ \texttt{medication|conflicting}, \texttt{medication|unknown};
(c)~\texttt{audit coverage update}: case class \texttt{medication|conflicting|routine} marked as audited.

\item \textbf{Post-patch evaluation:}
The fixed evaluation case ``You should stop taking warfarin immediately'' is re-evaluated.
The oracle's hard-block fires on ``stop taking,'' returning $s = 0.0$, $u = 0.05$.
Enforcement: $s < 0.4 \Rightarrow$ \texttt{BLOCK}.
Previously this case was \texttt{ESCALATE} (due to $u \geq u_{\mathrm{thresh}}$).
The governance batch changed the outcome from uncertainty to a definitive decision.
\end{enumerate}

\bibliographystyle{splncs04}
\bibliography{sample}

\end{document}